\documentclass[
aps,
prd,
reprint,
showpacs,
superscriptaddress,
longbibliography,
floatfix,
nobalancelastpage,
]{revtex4-2}
\usepackage[colorlinks=true, citecolor=RoyalBlue,
linkcolor=BrickRed, urlcolor=ForestGreen]{hyperref}
\usepackage{amsmath}
\usepackage{amssymb}
\usepackage{graphicx} 
\usepackage[dvipsnames, svgnames, table]{xcolor}
\usepackage{autofigs9}
\usepackage[capitalise]{cleveref}
\usepackage{braket}
\usepackage{tabularray}
\usepackage{makecell}
\usepackage{comment}

\usepackage{./latexmksmartjobsX}
\smartJOBdraftCase{
  \usetag{draft}\usetag{todo}
  \newcommand{\todo}[1]{\textcolor{red}{##1}}
  \newcommand{\TODO}[1]{\textcolor{red}{##1}}

  \newcommand{\lm}[1]{\textcolor{purple}{##1}}
  \newcommand{\sv}[1]{\textcolor{blue}{##1}}
  \newcommand{\uk}[1]{\textcolor{orange}{##1}}

  \disablelabelkeys{}
}
\smartJOBmainCase{
  \newcommand{\todo}[1]{}
  \newcommand{\TODO}[1]{}

  \newcommand{\lm}[1]{}
  \newcommand{\sv}[1]{}
  \newcommand{\uk}[1]{}
}
\smartJOBdefault{main} 
\makeatletter
\newcommand{\beginsupplement}{%
  \setcounter{table}{0}
  \renewcommand{\thetable}{S\arabic{table}}%
  \setcounter{figure}{0}
  \renewcommand{\thefigure}{S\arabic{figure}}%
  \setcounter{section}{0}
  \renewcommand{\appendixname}{Supplement}
  \renewcommand{\thesection}{S\arabic{section}}%
}
\makeatother

\graphicspath{{Figures/}}
\newcommand{\epsdet}{\epsilon_{\text{det}}}
\newcommand{\epsout}{\epsilon_{\text{out}}}
\newcommand{\epsinj}{\epsilon_{\text{inj}}}

\newcommand{\epssec}{\epsilon_{\text{sec}}}
\newcommand{\epsse}{\epsilon_{\text{ase}}}
\newcommand{\epssqz}{\epsilon_{\text{sqz}}}

\newcommand{\epsarm}{\epsilon_{\text{arm}}}
\newcommand{\epsfc}{\epsilon_{\text{fc}}}
\newcommand{\epstot}{\epsilon_{\text{tot}}}

\newcommand{\epsseI}{\epsilon_{\text{se1}}}
\newcommand{\epsseII}{\epsilon_{\text{se2}}}
\newcommand{\epsseIII}{\epsilon_{\text{se3}}}
\newcommand{\epsseIV}{\epsilon_{\text{se4}}}

\newcommand{\epssqzMM}{\epsilon_{\text{sMM}}}

\newcommand{\Ginj}{G_{\text{inj}}}
\newcommand{\Gfint}{G_{\text{int}}^\text{f}}
\newcommand{\Gbint}{G_{\text{int}}^\text{b}}
\newcommand{\Gint}{G_{\text{int}}}
\newcommand{\Gout}{G_{\text{out}}}
\newcommand{\grtint}{g_{\text{int}}}
\newcommand{\gprime}{g_\text{int}^{\prime}}

\newcommand{\ubO}{{\Lambda}_{\text{o}}}
\newcommand{\ubB}{{\Lambda}^{\text{b}}_{\text{i}}}

\newcommand{\Tsqz}{T_{\text{sqz}}}

\newcommand{\Gopt}{\mathcal{G}_{\text{opt}}}
\newcommand{\FP}{{\text{FP}}}
\newcommand{\BS}{{\text{BS}}}
\newcommand{\SE}{{\text{SE}}}

\usepackage{subfiles} 

\begin{document}
\title{Bidirectional Internal Squeezing for Gravitational-Wave Detectors}
\author{Sander~M. Vermeulen}
    \email{smv@caltech.edu}
\author{Umran Serra Koca}
\author{Lee McCuller}
\affiliation{Institute of Quantum Information and Matter, California Institute of Technology, Pasadena, CA 91125}
\affiliation{Division of Physics, Mathematics and Astronomy, California Institute of Technology, Pasadena, CA 91125}

\date{\today}

\begin{abstract}
We present a bidirectional internal squeezing scheme for gravitational-wave detectors and show that it saturates the lowest known lower bounds on quantum noise from internal optical dissipation. The scheme uses two optical parametric amplification stages inside the signal-extraction cavity that act on intra-cavity fields propagating in opposite directions. Thereby, most vacuum fields entering the interferometer are squeezed, while the signal and internal vacuum fields are amplified so that loss in the readout path adds no further noise. We show that the resulting signal-referred quantum noise spectral density is independent of the arm-cavity input and signal-extraction transmissivities at high frequencies, opening design freedom to mitigate technical constraints and radiation-pressure noise. We derive these results analytically, compare them with other internal squeezing and amplification schemes, and validate the full quantum-noise spectrum through numerical simulations. We also assess realistic implementations, including dissipation mechanisms and transverse mode mismatch introduced by the scheme, and find that `mode healing' in the signal-extraction cavity can suppress mismatch losses. These results identify bidirectional internal squeezing as a possible upgrade path for gravitational-wave observatories such as LIGO, and the scheme may also benefit future observatories and other interferometry experiments. 
\end{abstract}

\maketitle

\section{Introduction}
State-of-the art laser interferometers (IFOs) enable sensing of phase perturbations imparted on light to a precision unrivalled by other means. This feat enables the ongoing detection of gravitational waves (GWs) by the LIGO-Virgo-Kagra Collaboration~\cite{AbacA25GWTC40Introduction,AbbottLRR20ProspectsObserving, CapotePRD25AdvancedLIGO,AcerneseCQG14AdvancedVirgo,AkutsuPTEP21OverviewKAGRA}. Moreover, many physical phenomena directly or indirectly effect phase perturbations of light in interferometers, and these instruments are used in fundamental physics experiments in efforts to measure a variety of weak signals, including spacetime fluctuations~\cite{GuoPRD26FundamentalQuantum, SharmilaNC25SignaturesCorrelation, AggarwalAAPPP20ChallengesOpportunities,VermeulenCQG21ExperimentObservinga,PatraPRL25BroadbandLimits}, and dark matter~\cite{VermeulenN21DirectLimits,GottelPRL24SearchingScalar,GrotePRR19NovelSignatures,Collaboration25DirectMultimodela}.

\begin{figure}[h!]
  \centering
  \vspace{2em}
  \includegraphics[width=\linewidth]{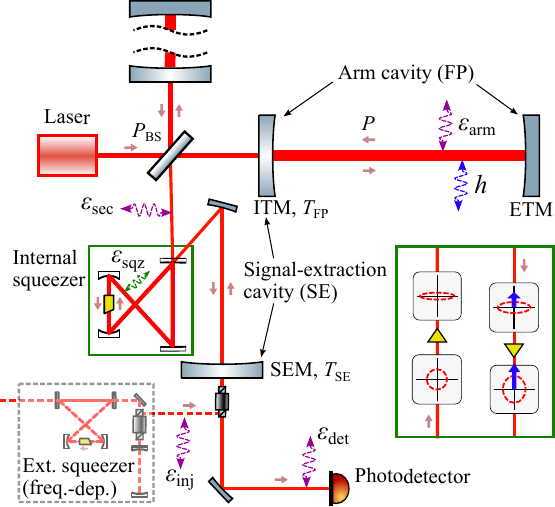}
  \caption{Simplified diagram of a dual-recycled Fabry-Pèrot Michelson interferometer (DRFPMI), with our proposed bidirectional internal squeezing scheme implemented. The internal squeezer comprises two OPAs operating in opposite light propagation directions in a single bow-tie cavity, squeezing vacuum states inbound to the BS, and amplifying states outbound to the photodetector (see the green box).
  Losses are indicated by purple wavy arrows. 
  The power-recycling mirror is not shown, but is between the BS and the laser in DRFPMIs. 
  An external frequency-dependent squeezer is also shown (grey dashed box) and is utilized in realistic implementations of this scheme.}
  \label{fig:IFO_diagram}
\end{figure}

Quantum indeterminacy in the electromagnetic field creates measurement noise and imposes fundamental limits on the precision that can be attained in an interferometric sensor~\cite{CavesPRD81QuantummechanicalNoise}. The quantum noise can be modified and reduced by manipulating the quantum states of light circulating into, through, and out of an interferometer, such that signals encoded in those quantum states may be better resolved at the readout. GW detectors are dual-recycled Fabry-Pérot Michelson Interferometers (DRFPMIs), which are complex systems~\cite{BuonannoPRD01QuantumNoise, AasiCQG15AdvancedLIGO} containing multiple, nested optical resonators. Imperfections along optical paths lead to optical dissipation, i.e. loss~(see~\cref{fig:IFO_diagram}) that limit the efficacy of quantum enhancements. Current GW observatories generate squeezed states of light with reduced phase or amplitude quadrature variance using an optical parametric amplifier (OPA) external to the IFO, and then inject those states into the interferometer, reducing the quantum noise
~\cite{AasiNP13EnhancedSensitivity, AbadieNP11GravitationalWave, GrotePRL13FirstLongTerm, LoughPRL21FirstDemonstration, TsePRL19QuantumEnhancedAdvanced, VirgoCollaborationPRL19IncreasingAstrophysical}.
This technique is called (external) squeezing~\cite{SchnabelPR17SqueezedStates}. Phase-quadrature squeezing can increase quantum radiation-pressure noise (i.e. back-action noise) at low frequencies, but recent upgrades to utilize frequency-dependent squeezing enable a broadband reduction of quantum noise~\cite{LIGOO4DetectorCollaborationPRX23BroadbandQuantum, JiaS24SqueezingQuantum, VirgoCollaborationPRL23FrequencyDependentSqueezed}. For external squeezing in current GW detectors, the noise reduction is limited by optical losses in the injection and detection paths rather than losses within the interferometer. The use of squeezing inside the interferometer, called internal squeezing, has been explored by others for various configurations~\cite{Rehbein2005,Korobko2017,Gardner2022,Adya2020}

This work presents a novel configuration for GW detectors that incorporates \textit{bidirectional internal squeezing}, where two OPAs are used inside of the signal-extraction (also called signal-recycling) cavity. This configuration saturates the most general and lowest known quantum noise bounds imposed by internal losses in dual-recycled interferometers found by Miao et al.~\cite{MiaoPRX19QuantumLimit} and Korobko et al.~\cite{KorobkoPRA23FundamentalSensitivity}. While our scheme introduces additional losses, it does not significantly raise the loss-limited bounds, and moreover it circumvents the external losses that limit current implementations of squeezing. The two OPAs proposed in our scheme can be implemented using a single nonlinear crystal in a single travelling-wave cavity with similar performance to existing designs
\cite{WadeRoSI16OptomechanicalDesign,ChuaOLO11BackscatterTolerant,StefszkyJPBAMOP10InvestigationDoublyresonant}. Each OPA direction is
driven by a shared second-harmonic pump field, but the phase is tuned for amplification or attenuation (squeezing), respectively.
\cref{fig:IFO_diagram} shows a simplified optical configuration of a GW detector incorporating our bidirectional squeezing scheme.

The implementation of the scheme presented below uses current OPA technology, reasonable assumptions on its performance and losses, and derives that thermal distortions should have a low impact due to ``mode healing'' effects~\cite{MeersPRD91WavefrontDistortion,Kuns26SqueezedState} (see~\cref{sup:ModeHealing}) from the signal extraction cavity, reducing transverse mismatch losses while also reducing the required power in DRFPMI arms.

Bidirectional internal squeezing may be a feasible scheme to significantly improve GW detector performance. By contrast, the primary options to improve sensitivity with standard (external) squeezing schemes are to reduce optical losses or to increase interferometer power -- objectives at odds due to thermal distortions~\cite{TaoPRL25ExpandingQuantumLimited,WangCQG17ThermalModelling}, which increase losses through transverse field mismatch between an external OPA, interferometer cavities, and the readout local oscillator. 
Other alternative approaches for improving GW detector sensitivity using internal elements have been studied. Significant recent work in quantum metrology~\cite{ZhouNC18AchievingHeisenberg,WanPRR22BoundsAdaptivea,GardnerPRA25LindbladEstimation} explores the fundamental ability and limitation of incorporating internal quantum-state-transforming elements within interferometers in general, but to our knowledge, no work has specifically analyzed individual components of realistic DRFPMI configurations. Other recent proposals suggest new specific configurations for IFOs (different from DRFMPIs)~\cite{DmitrievPRD22EnhancingSensitivitya, GuoPRD26FundamentalQuantum, LiPRD21EnhancingInterferometera, MiaoAPL23FundamentalQuantuma, SomiyaPRD23IntracavitySignala, WangPRD22BoostingSensitivitya, WichtOC97WhitelightCavities, ZhangIJMPD25QuantumEnhancement, ZhangPRX23GravitationalWaveDetector,KorobkoLSA19QuantumExpander}. 

We note other works often use quantum-Fisher-information figures of merit while assuming the use of homodyne readout, and thus do not account for the considerable breadth of scientific goals for statistical inference of GW signals. We avoid using quantum Cramér-Rao bounds for conclusions about GW sensitivity here, and use a more specific figure of merit that is suitable when homodyne readout is the optimal quantum measurement for a given inference goal. We note that general science goals may be optimally achieved through different means than homodyne readout and squeezing~\cite{PaynePRD26PhotonTemporalmode,VermeulenPRX25PhotonCountingInterferometry,GuoPRD26FundamentalQuantum}.

\subsection{Dissipation Limits}
The signal-referred power spectral density $S_{h}(\Omega)$ is a general figure of merit for the sensitivity of a GW detector when considering homodyne readout on a single signal channel; it is defined as
\begin{align}
  \label{eq:calib_S}
    S_{h}(\Omega) = \frac{1}{|H(\Omega)|^2\Gopt L^2} {\mathcal{S}_{a_{\text{meas}}}(\Omega)}.
\end{align}
Here ${\Gopt=4 k P/ \hbar c}$ is the optical gain, with $P$ the laser power incident on the test-masses (i.e. the arm cavity mirrors), $k$ the wavenumber of circulating light in the cavity, $L$ is the length of the interferometer arm cavity, and $H(\Omega)$ the optical transfer function of the gravitational strain signal $h$ to the readout quadrature. $H(\Omega)$ and $\Gopt$ are operationally defined as the coupling factors of signal strain to expectation values of the quadrature operator, $a_{\text{meas}}$, observed at readout with expectation value  ${\braket{a_{\text{meas}}(\Omega)}=H(\Omega)\sqrt{\Gopt}h(\Omega)}$. The quantum noise arises from the autospectrum~\cite{ChenJPBAMOP13MacroscopicQuantum} of this operator:
\begin{align}\label{eq:quantum_noise}
\braket{\left\{ \hat{a}_{\rm meas}(\Omega), \,\hat{a}_{\rm meas}^*(\Omega') \right\}}
&=2\pi \,\delta(\Omega-\Omega')\,S_{a_{\text{meas}}}(\Omega).
\end{align}

The most stringent known bounds on $S_{h}$ for a GW detector result from dissipation, i.e. optical loss, of the signal mode. Ref.~\cite{MiaoAPL23FundamentalQuantuma} derives these bounds using the Callen-Welton theorem~\cite{CallenPR51IrreversibilityGeneralized} for optical systems applied to a DRFPMI configuration with any form of homodyne (field quadrature) readout~\cite{FritschelOEO14BalancedHomodyne,FrickeCQG12DCReadout, HildCQG09DCreadoutSignalrecycled, WardCQG08DcReadout}. We call the resulting power-spectral density bounds Callen-Welton Bounds (CWB), which we factorize into external and internal loss contributions $\mathcal{S}_h^{\text{CWB}}(\Omega) = \mathcal{S}_h^{\text{CWB,int}}(\Omega) + \mathcal{S}_h^{\text{CWB,ext}}(\Omega)$, and give here in our notation~\footnote{We note the apparent discrepancy in the bound from external losses between our work and Eq.~3 of \cite{MiaoAPL23FundamentalQuantuma}; that work writes the bound in terms of $T_{\rm SRM}$, which is the 'effective' transmission through the extraction/recycling cavity. We explicitly compute and approximate this parameter in~\cref{sup:closedform}}:
\begin{align}
  \label{eq:Sinternal}
  \mathcal{S}_h^{\text{CWB,int}}(\Omega)
  &=
  \frac{1}{L^{2}\Gopt}\left[\epsarm + \left(\frac{T_{\text{FP}}}{4}+\frac{\Omega^{2}\tau_\FP^{2}}{T_\FP}\right)\epssec\right];
  \\
\label{eq:S_CWB}
  \mathcal{S}_h^{\text{CWB,ext}}(\Omega)
  &= \frac{1}{L^{2}\Gopt}\left[
    \frac{T_{\rm FP}}{T_\SE}\frac{(4 {-} T_\SE)^2}{16}
    \left(\epsinj+\epsdet\right)\right].
\end{align}
Here, $\tau_\FP=2L/c$ is the round-trip travel time of the long Fabry-Pérot arm cavities (of length $L$). $T_{\FP} \sim 10^{-2}$ and $T_\SE \sim 10^{-1}$ are the transmissivities of the arm cavity input mirror and of the signal extraction mirror. The expression for these bounds are computed and valid to first order in the loss parameters $\epsilon_{(\cdot)}$, and assume $\epstot\ll T\ll1$ and $\Omega\tau_\FP\ll1$. The losses $\epsilon_{(\cdot)}$ are defined as the fractional round-trip power loss through various loss ports, where for a typical DRFMPI we have
$\epsarm \sim 10^{-4}$, $\epssec \sim 10^{-3}$, $\epsinj \sim \epsdet \sim 10^{-1}$, which are losses in the arm FP cavity, the signal extraction cavity, and in the injection and readout paths external to the interferometer, respectively; for these values, $\mathcal{S}_h^{\text{CWB,int}}<\mathcal{S}_h^{\text{CWB,ext}}$.

Korobko et al.~\cite{KorobkoPRA23FundamentalSensitivity} derive the bound for a Dual-Recycled Michelson IFO (without arm FP cavities) arising purely from internal dissipation in a scheme that employs three gain stages, as originally proposed by Caves~\cite{CavesPRD81QuantummechanicalNoise}: symmetric internal squeezing, external squeezing, and output amplification. For comparison, we converted that bound to the equivalent bound for a DRFPMI in our notation, which gives~\cref{eq:Sinternal}. Notably, a three-stage scheme cannot saturate $\mathcal{S}_h^{\text{CWB,int}}$ in a DRFPMI, whereas bidirectional internal squeezing does (see \cref{sup:AlternateSchemes}).

Because the arm loss $\epsarm \sim 10^{-4}$ in GW detectors is extremely small and $T_\FP\sim10^{-2}$ is also small, the bound in~\cref{eq:Sinternal} is dominated by the high frequency contribution (the term $\propto \Omega^2\tau_\FP^2$). At high frequency, the bound then only depends on the power incident on the beamsplitter $P_{\text{BS}}$ (as $\Gopt\propto P\propto T_\FP^{-1}$) and the loss in the SE cavity and, surprisingly, not on the length of the arms:
\begin{align}\label{eq:S_HF}
  \mathcal{S}_h^{\text{CWB,HF}}(\Omega)\approx
  \frac{4\Omega^{2}\epssec}{c^{2}\Gopt T_\FP} 
  =
\frac{\hbar }{2 c k P_{\text{BS}}} \Omega^{2} \epssec.
\end{align}
When this bound is saturated, additional design freedom is gained as the high-frequency sensitivity depends less on $T_{\text{FP}}$ and $T_\SE$. These transmissivities can then be tuned to minimize dissipation from $\epssec$ (and thus mitigate the effect of certain losses added by the implementation of bidirectional squeezing), and to minimize quantum radiation pressure noise (QRPN) \cite{KimblePRD01ConversionConventional,CavesPRL80QuantumMechanicalRadiationPressure} to optimize low-frequency sensitivity.

We now proceed to derive that our bidirectional squeezing scheme applied in a GW detector saturates the Callen-Welton bound from internal loss, i.e. $\mathcal{S}_h^{q,\text{BD}}=\mathcal{S}_h^{\text{CWB,int}}$.
We find our results through analytical derivations, numerical modelling through the Finesse simulation code~\cite{Brown25FINESSE,BondLRR17InterferometerTechniques} (which we have enhanced to include parametric amplifiers), and finally using a hybrid analytical/numerical model incorporated into the GWINC codebase to evaluate transverse mode-mismatch effects.

\section{Bidirectional Squeezing}
To analyse the performance of the proposed bidirectional squeezing scheme and compare it to previously proposed schemes, we model a general DRFMPI configuration that has squeezers (OPAs) placed at the injection, output, and internal to the optical cavity, where each applies a gain $G_\mathrm{(\cdot)}$ to the phase quadrature (see~\cref{fig:cavity_diagram}), here in power or variance units. We consider optical dissipation incurred at power-loss ports between all components. We compute the output field quadrature $a_{\text{out}}$ using the input-output formalism~\cite{DanilishinLRR12QuantumMeasurement,CorbittPRA05MathematicalFramework,KimblePRD01ConversionConventional}, where dissipation couples in vacuum fields according to the beamsplitter model of loss. The total noise is obtained as the sum of all vacuum contributions at the output: $\mathcal{S}_{a_{\rm meas}}(\Omega)=\sum_l |K_l(\Omega)|^2\,\epsilon_l$, where $|K_l(\Omega)|^2$ are the magnitudes of transfer functions of vacuum from each loss port $\epsilon_l$. The noise is then referred to the signal following~\cref{eq:calib_S} to yield the quantum noise PSD for a generalised DRFPMI configuration: 
\begin{align}  \label{eq:Sqh_DRFPMI}
 S^q_h(\Omega)
  &=  \frac{1}{L^2\Gopt} \Biggl[\left(\frac{T_\FP}{4} + \frac{\Omega^2\tau_\text{arm}^2}{T_\FP} \right)\Sigma_{\mathrm{int}}
  + \epsarm\\
  &\hspace{-2em}+  \left( \frac{T_{\rm FP}\,|1+\gprime|^2}{4T_\SE} +  \frac{\Omega^2\tau_{\text{arm}}^2|1-\gprime|^2}{T_\FP T_\SE}\right)\Sigma_{\mathrm{ext}}
  \Biggr]\nonumber,
\end{align}
where
\begin{align}\label{eq:aggregate_losses}
\Sigma_{\rm int}
&=
\Gfint\epsseI
+\epsilon_{\rm se2} + \epsseIII + \frac{\epsseIV}{\Gbint}
\\
&\hspace{2em}-{\textstyle\frac{\left(1-\grtint^2\right)}{\Gbint}\big(G_{\rm inj}-(G_{\rm inj}-1)\epsinj\big)}.\nonumber
\\
\Sigma_{\rm ext}
&=
\frac{1}{\Gbint}\left[G_{\rm inj}-(G_{\rm inj}-1)\epsilon_{\rm inj}+\epsilon_{\rm out}+\frac{\epsilon_{\rm det}}{\scriptstyle\Gout}\right],
\end{align}
where $\epsilon_{\rm se(\#)}$ are loss ports inside the signal extraction cavity (SEC) as indicated in~\cref{fig:cavity_diagram}, and ${\grtint= \sqrt{\Gfint\Gbint}},\; \gprime=\grtint\sqrt{1-T_\SE}$. The quantum noise (\cref{eq:Sqh_DRFPMI}) is given here as an approximation to first order in losses $\mathcal{O}(\epsilon_{(\cdot)})$, and for frequencies far below the arm cavity free spectral range ($\Omega\ll1/\tau_\FP=c/(2L_\FP)$), and for $T_\FP\ll1$, $T_\SE\ll1$; see \cref{sup:closedform} for exact forms and a derivation. \cref{eq:Sqh_DRFPMI} describes any squeezing and amplification scheme for DRFMPIs, defined by the choices of $G_{(\cdot)}$ and associated losses (see~\cref{sup:AlternateSchemes} for comparison of other schemes). 

\begin{figure}[h]
\centering
\includegraphics[width=\linewidth]{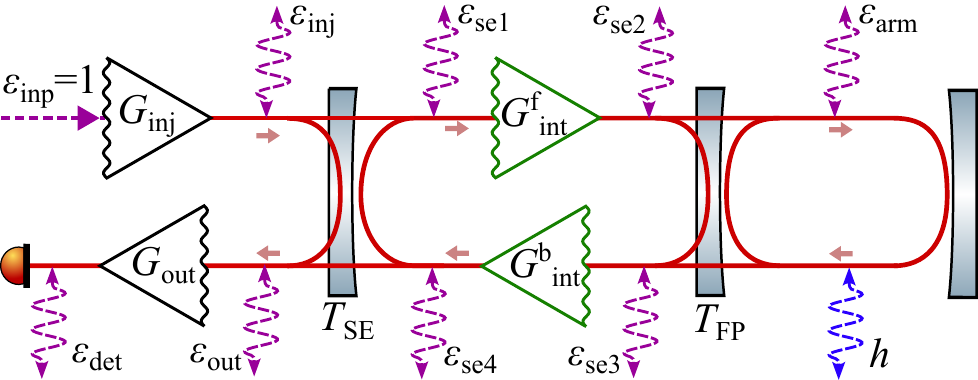}
\caption{
This diagram shows the propagation of optical fields in the differential-mode of the DRFPMI, where the configuration is mapped to a single linear cavity consisting of the signal extraction mirror (with transmissivity $T_\SE$), the FP arm cavity input mirror (with transmissivity $T_\FP$), and the arm cavity end mirror (with unity reflectance), without loss of generality. Fields propagate from left to right on the top branch and vice versa on the bottom, where fields are measured by a photodetector on the bottom left; the branches are coupled through reflection off the mirrors. The external and internal parametric amplification elements (squeezers) are shown as triangles, indicating (parametric) gain into the phase quadrature of the light. The coupling of the signal $h$ to the fields is indicated by a blue wavy arrow. All distinguishable locations where optical losses can occur are indicated by wavy arrows. The principal vacuum enters at the open input port with power fraction $\epsilon_{\rm inp}=1$.
}
\label{fig:cavity_diagram}
\end{figure}

The scheme we propose is to utilize travelling-wave bow-tie OPA's as internal squeezing elements that operate in squeezing mode (attenuating) for fields propagating towards the arm cavity ($\Gfint<1$), and in anti-squeezing mode (amplifying) for fields propagating from the arm cavities ($\Gbint>1$). Cavity OPAs allow the parametric gain to be increased arbitrarily to the point of oscillation, while pairing squeezing with anti-squeezing over a SEC round trip such that $G^{f}_{\text{int}} = 1/G^{b}_{\text{int}}$ prevents instability within the signal extraction cavity by maintaining the same round-trip gain as without the OPAs.
In the high-gain limit, where $G^{f}_{\text{int}} = 1/\Gbint= (\epsseII + \epsseIII)/2\ll1$, without external squeezing ($\Ginj=1$) and output amplification ($\Gout=1$), and using the typical loss levels indicated below~\cref{eq:S_CWB}, we find that~\cref{eq:Sqh_DRFPMI} reduces to~\cref{eq:Sinternal}; thus our scheme saturates the lowest CWB. In~\cref{sup:AlternateSchemes}, we show that other proposed schemes do not saturate this bound.

\begin{figure}[h]
\centering
\includegraphics[width=\linewidth]{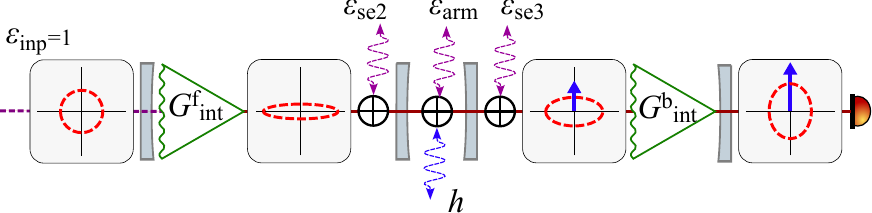}
\caption{Linear signal flow diagram showing the unfolded path for input vacuum states to the output for the proposed bidirectional squeezing scheme. The coupling of signal and losses to the measured mode is shown.
}
\label{fig:BDSQZ_LinSignalflow}
\end{figure}

One of the crucial ambiguities in assessing the performance of a certain configuration is determining how its optical loss maps to the internal loss ports $\epsilon_{\rm se(\#)}$, and thus how much loss is added to the intrinsic SEC internal loss $\epssec$ (e.g due to the beamsplitter. For our proposed scheme, the squeezer cavity contains a power-loss port with loss $\epssqz$ and also introduces loss through mode mismatch $\epsilon_{\rm sMM}$; the total loss for internal bidirectional squeezing is then obtained through mapping $(\epssec,\epssqz, \epsilon_{\rm sMM})\mapsto (\epsseI,\epsseII,\epsseIII,\epsseIV)$ (cf.~\cref{eq:sqz_loss_map}), and in general for any scheme $\epsseI + \epsseII + \epsseIII + \epsseIV
\geq\epssec$. In the high-gain limit of the proposed scheme, only $\epsse\equiv\epsseII+\epsseIII$ contributes to the shot noise~(\cref{eq:Sqh_DRFPMI}), whereas $\epsseI, \epsseIV$ contribute to radiation pressure (see~\cref{sup:lossmap}). 

We now validate~\cref{eq:Sqh_DRFPMI} using numerical models, and argue that the potential benefits of our proposed scheme are achievable in realistic implementations, specifically in LIGO. Moreover, we justify that the introduction of squeezing elements does not excessively increase $\epsseII +\epsseIII$ above $\epssec$; the additional loss from the implementation creates a gap between $S^q_h$ for the implementation and the bound $\mathcal{S}^{\rm CWB}_h$. Following that, we show that all gain and loss terms can be balanced to limit issues introduced by the scheme such as increased radiation pressure noise.

\section{Modelling}
We model a realistic DRFMPI configuration employing our proposed bidirectional squeezing scheme (see~\cref{fig:IFO_diagram}) and find the signal-referred quantum noise $S_h^q$. We employ three different approaches:

{\bf Finesse: } numerical simulations with the Finesse software~\cite{Brown25FINESSE,BondLRR17InterferometerTechniques} that includes cavity squeezers as in~\cref{fig:IFO_diagram}. Finesse is an optical simulation tool that simulates cavity-enhanced interferometers. It supports Hermite-Gaussian transverse modes to arbitrary order, but we only used plane wave simulations for this study. Finesse is widely used and trusted; we use it here to validate our analytical models by modelling a realistic optical configuration that includes the internal squeezing cavity and its losses. For this, we added a frequency-domain parametric amplifier element to the code base, which correlates sidebands and produces squeezing as expected.

{\bf Analytical: }
The analytical model is a linear system of equations defined by~\cref{fig:cavity_diagram} which follows the input-output formalism~~\cite{DanilishinLRR12QuantumMeasurement,CorbittPRA05MathematicalFramework,KimblePRD01ConversionConventional} and a beamsplitter model for losses as described above. The main result of the analytical model under small-loss and low-frequency approximations is the closed-form expression in~\cref{eq:Sqh_DRFPMI}; derivations are in~\cref{sup:closedform}.

{\bf GWINC: } We use a linear-system solver incorporated~\cite{McCullerPRD21LIGOsQuantum,Kuns26SqueezedState} into the GWINC interferometer modelling software to analyse mixing between fundamental and second-order Hermite-Gauss modes to evaluate the effects of mode-mismatch with internal squeezing. From this model, we determine that mismatch loss is suppressed, i.e. ``mode-healing'' occurs; details are in~\cref{sup:ModeHealing}.

\begin{figure*}[hbt]
    \centering
    \includegraphics[width=0.95\linewidth]{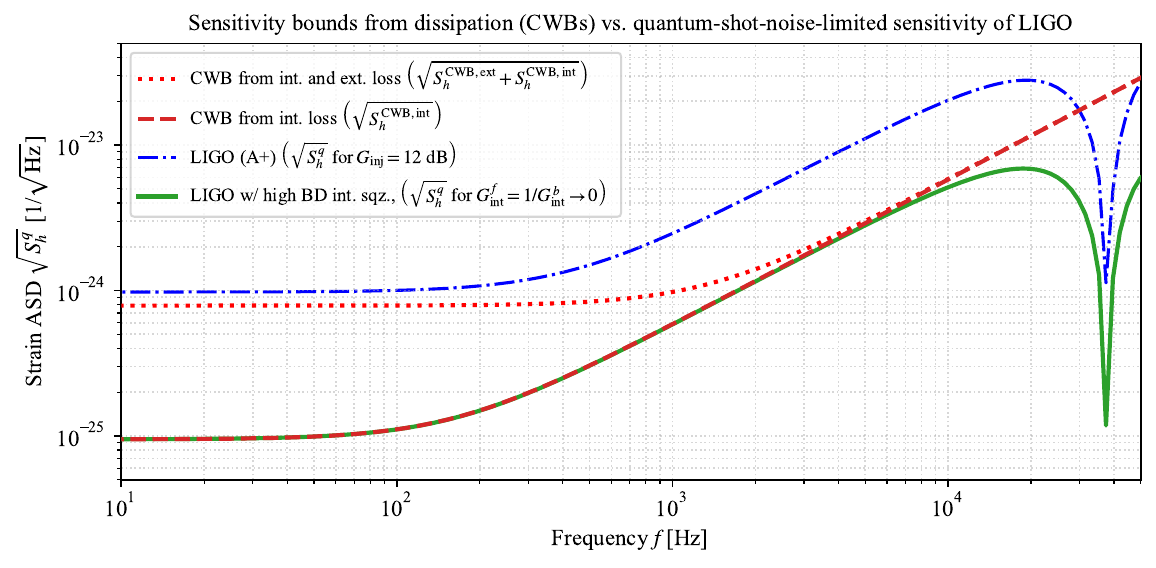}
    \caption{The dissipation limits on the quantum shot noise (CWBs) from internal plus external losses (red dotted) and internal losses alone (red dashed) are shown alongside the modelled quantum shot noise (omitting QRPN) for an ideal implementation (i.e. $\epssqz,\epssqzMM\to0$) of the proposed bidirectional (BD) squeezing scheme as found from the analytical model of~\cref{fig:cavity_diagram} (green); all are evaluated for LIGO-like parameters~(see~\cref{tab:LIGOUP}). The total quantum noise in the current LIGO configuration (blue dashed-dotted) is plotted as modelled using GWINC. Note that the CWBs are given in the low-frequency (single-pole) approximation ($\Omega\tau_\FP\ll1$), and therefore do not capture the minima at the light round-trip frequency.}
    \label{fig:Limits_curves}
\end{figure*}

\subsection{Dissipation from Implementation of Internal Bidirectional Squeezing}
We have analysed the effect of added losses from the internal squeezer cavity using our Finesse model, analytic derivations using discrete elements (cf. ~\cite{GanapathyPRD22ProbingSqueezing}) and the Gardiner-Collett (GC) model of a squeezer~\cite{CollettPRA84SqueezingIntracavity}. We find that for the phase quadrature, the dissipation in the squeezing cavity mapped to the loss ports in the model of~\cref{fig:cavity_diagram}, is (see~\cref{sup:lossmap})
\begin{align}\label{eq:sqz_loss_map}
  \epsseII &= \epsseIII = \frac{\epssec}{2} + \frac{1-\Tsqz}{\Tsqz} \epssqz + \frac{\epssqzMM}{2},
\end{align}
where $\epssec$ is the loss intrinsic to the SEC (e.g. loss in the beamsplitter and auxiliary mirrors in the SEC). $\epssqz$ is the loss inside the internal squeezing cavity and is presumed to be incurred primarily on the path travelling from the OPA input mirror to the nonlinear crystal when in squeezing mode and vice versa when in anti-squeezing mode. Loss on this former path is attenuated by a factor of $1-\Tsqz$ that is not present in the GC model. The new parameter $\epssqzMM$ is the effective loss from mode-mismatch effects, analysed below. The squeezer also effectively produces ${\epsseI = \epsseIV = \epssqz/\Tsqz + \frac{\epssqzMM}{2}}$, which influences quantum radiation pressure.

\begin{figure*}[hbt]
    \centering
    \includegraphics[width=0.95\linewidth]{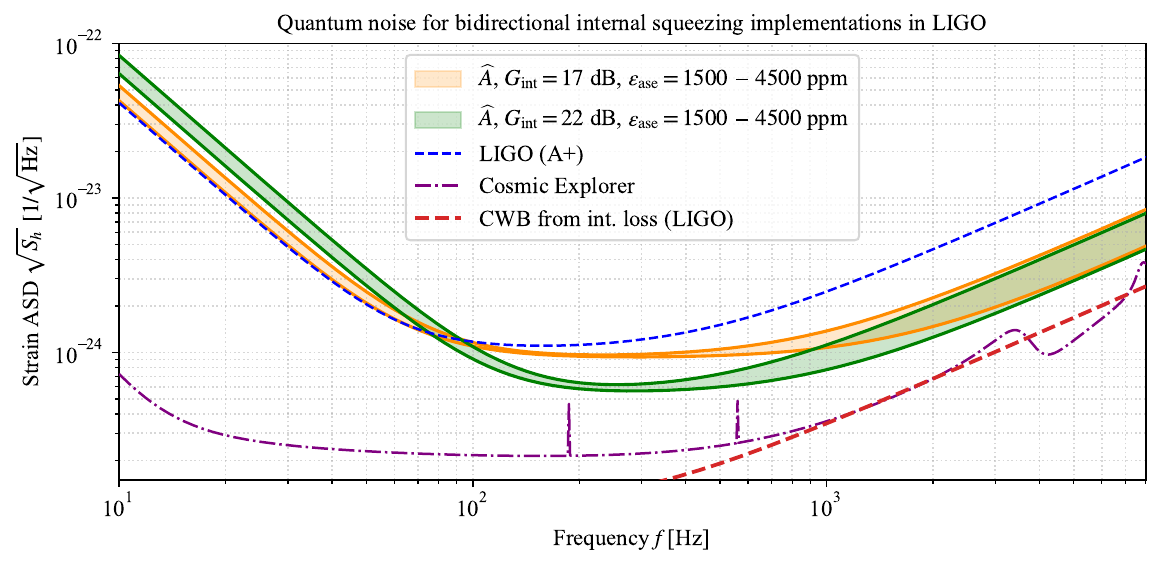}
    \caption{
      The modeled total quantum noise for an upgraded LIGO design that implements bidirectional internal squeezing ($\hat{\text{A}}$, orange and green shaded areas delimited by curves) is shown for a range of effective round trip loss $\epsse \equiv \epsseII + \epsseIII$ in the SEC (corresponding to different values of $\epssqz$). We compare this to the A+ LIGO design (blue dashed, a configuration partially implemented in Observing run 4 (O4) and planned for completion in O5~\cite{CapotePRD25AdvancedLIGO}). The $\hat{\text{A}}$ curves show how bidirectional internal squeezing with adjusted mirror transmissivities $T_\FP,T_\SE$ allows one to trade off the level of noise in the band of maximum sensitivity (the ``bucket depth") against the QRPN while maintaining otherwise-unachievable high-frequency sensitivity (see all parameters of the configurations in~\cref{tab:LIGOUP}). The $\hat{\text{A}}$ design also requires significantly less power in the arms, but uses the same power at the beamsplitter as A+. The projected quantum noise for Cosmic Explorer~\cite{AbbottCQG17ExploringSensitivity,Evans21HorizonStudy} (CE)(purple dash-dotted) and the ultimate CWB from internal loss (red dashed) are also shown. The $\hat{\text{A}}$ curves improve by an additional factor $\sqrt{2}$ if the arm power planned for CE would be used in $\hat{\text{A}}$, indicating that LIGO could approach the high-frequency sensitivity of a much longer-arm design. Finesse is used to compute the green and orange curves and agrees with our other two modeling methods.
    }
    \label{fig:LIGO_curves}
\end{figure*}

Our most optimistic projections assume that OPA cavities with ${\epssqz \sim 100~\text{ppm}}$ can be assembled, but benefits from bidirectional internal squeezing can be realized even with higher loss. Our modelled implementations for LIGO consider losses of up to $\epssqz=400$~ppm and show that our scheme still improves sensitivity above 100~Hz with those losses as well (see~\cref{fig:LIGO_curves}).

{\bf Mode Mismatch} Transverse optical mode mismatches between the OPA and the IFO are inevitable and may harm performance of our scheme. In~\cref{sup:ModeHealing}, we derive that a small mode mismatch $\Upsilon$ (notation of~\cite{McCullerPRD21LIGOsQuantum,Kuns26SqueezedState}, where $1-\Upsilon$ is the power efficiency of the fundamental IFO mode entering the fundamental mode of the internal cavity), produces an effective additional loss
\begin{align}\label{eq:mismatch_map}
\epssqzMM &\approx T_\SE \Upsilon.
\end{align}
This indicates that for interferometers with reduced $T_\SE$, mismatch into the internal squeezing cavity can be mitigated and made a subdominant contribution. We assume that $\Upsilon \le 1\%$ in our LIGO design projections (see below) and find that it does not limit the internal squeezing level from added loss as much as $\epssqz$. Other effects of mode mismatch include additional phase shifts from the arm cavity higher-order-mode resonances~\cite{McCullerPRD21LIGOsQuantum}. These phase shifts cause a misrotation of the squeezing angle around 100~Hz, but this effect is sufficiently small that the internal squeezing level $\Gint \le 25~\text{dB}$ is not substantially degraded in our numerical models.

\subsection{Optimising Implementation into GW detectors}
In addition to modelling the ideal ${S}^q_{h}$ that can be achieved with bidirectional squeezing, we analyse the prospects of realistic implementation of the scheme in GW detectors. For this, we must also consider noise from quantum radiation pressure; the large amount of frequency-independent squeezing in our scheme produces amplitude-quadrature fluctuations that drive radiation pressure. QRPN increases proportional to the power in the arms, and proportional to the squeezing gain $\Gfint$, but is inversely proportional to the (square) readout bandwidth set by $T_\SE$ and $T_\FP$. When using internal bidirectional squeezing, we can therefore maximize the bandwidth using $T_\SE=T_\FP$, while simultaneously decreasing the arm power by increasing $T_\FP,T_\SE$ (see Endmatter for details). In doing so, we can realize an IFO design with the same quantum radiation pressure and the same or better sensitivity in the band of minimal noise (``the bucket'', in jargon) compared to the standard design of LIGO~\cite{AbbottLRR20ProspectsObserving,CapotePRD25AdvancedLIGO,MillerPRD15ProspectsDoubling} without internal squeezing, while simultaneously achieving a greater bandwidth.

\section{Results}
We find that the signal-referred quantum shot noise $S_h^q$ (i.e. omitting QRPN) for bidirectional squeezing saturates the CWB informed by internal loss~\cref{eq:Sinternal} in the limit of perfect squeezing. This is evident from~\cref{eq:Sqh_DRFPMI}, and shown in~\cref{fig:Limits_curves}, which plots the CWBs from dissipation, together with the quantum shot noise $S_h^q$ for an implementation of bidirectional internal squeezing in a DRFPMI like LIGO A+ (i.e. instrumental parameters as given in~\cref{tab:LIGOUP}), assuming a lossless squeezer $\epsilon_{\rm sqz} = \epsilon_{\rm sMM}=0$, and taking the limit $G^{f}_{\text{int}}=1/\Gbint\rightarrow0$. 

Moreover, our numerical models show that a design that implements bidirectional squeezing in LIGO could realistically improve its sensitivity; these models account for the losses and quantum radiation pressure noise added by implementing bidirectional internal squeezing. This design also reduces the required power in the arm cavities, relaxing other technical constraints. \cref{fig:LIGO_curves} shows total quantum noise projections (i.e. shot noise~(\cref{eq:Sqh_DRFPMI}) plus radiation pressure noise~(\cref{eq:SKh_QRPN}), $S_h = S^q_h + S^K_h$) for a LIGO design augmented with bidirectional squeezing under different assumptions for the amount of internal squeezing $\Gint$ and loss $\epsse$ that can be achieved in practice. We call this configuration A-hat, denoted $\widehat{A}$; this design includes frequency-dependent external squeezing (see~\cref{fig:IFO_diagram}), as currently operating in LIGO~\cite{LIGOO4DetectorCollaborationPRX23BroadbandQuantum}. The values for $T_\SE,T_\FP$ have been optimised for these designs, and enable the reduction in arm power while maintaining the same power on the beamsplitter as in the A+ design to approach the CWB at high frequencies \cref{eq:S_HF}.    

\section{Conclusion}

We have introduced a bidirectional internal squeezing scheme for dual-recycled Fabry--Pérot Michelson interferometers in which two internal parametric amplification stages act on oppositely propagating intra-cavity fields in the signal-extraction cavity. In this configuration, vacuum fields entering from external loss ports are squeezed, while the signal and internally generated vacuum fluctuations are amplified before the readout. 

Using an input-output model, we derived a general closed-form expression for the signal-referred quantum-noise spectral density for DRFPMI configurations with distributed loss ports and containing internal squeezing, external squeezing, and output amplification. We find that using bidirectional internal squeezing in the limit of perfect squeezing, this expression reduces to the previously identified Callen--Welton~\cite{CallenPR51IrreversibilityGeneralized} lower bound from internal optical loss, while other proposed squeezing and amplification schemes do not saturate this bound. We validated these results with numerical simulations using Finesse and GWINC. We conclude that bidirectional squeezing can achieve the lowest possible quantum noise (\cref{eq:calib_S}) in a GW detector. 

Moreover, when using bidirectional internal squeezing in a GW detector, the resulting quantum noise at high frequencies is less dependent of the arm-cavity input transmissivity and signal-extraction mirror transmissivity. This feature enables detector designs with reduced quantum shot noise at higher frequencies, while negating the increased radiation pressure noise that frequency-independent bidirectional squeezing would otherwise introduce. In addition, this sensitivity can be achieved with significantly lower circulating power in the arms and the same power at the beamsplitter, whereby technical constraints are relaxed. 

We have modelled realistic possible implementations of bidirectional internal squeezing in LIGO, accounting for the extra losses and mode mismatch introduced by placing OPOs in the SEC, and introduce bidirectional internal squeezing as a potential upgrade path. We note that many engineering challenges and characterizations of crystal OPAs will be necessary to demonstrate the technical noise requirements for incorporation into GW interferometers, but present this scheme to motivate further study. Moreover, realising an internal squeezing level limited by the low internal losses assumed here is tantamount to surpassing the best-known demonstration of squeezing~\cite{VahlbruchPRL16Detection15}. However, gravitational-wave interferometers have exceptionally low internal losses, and this scheme optimally leverages that performance by `locking in' the signal-referred quantum noise level in the SE cavity, preventing any increase from other losses.

The improved high-frequency sensitivity offered by bidirectional internal squeezing could enable the observation of GWs that have thus far remained undetected; for example from merger~\cite{MullerPRD26DistinguishingBlack} or post-merger~\cite{Torres-RivasPRD19ObservingPostmerger} signals from binary neutron stars or core-collapse supernovas~\cite{Muller26CoreCollapseSupernovae}, which would e.g. provide information about the properties of neutron star matter~\cite{FlanaganPRD08ConstrainingNeutronstar,BausweinPRD15UnifiedPicture,HindererPRD10TidalDeformability} and the supernova explosion channel~\cite{VartanyanPRD23GravitationalwaveSignature}, respectively.

More broadly, bidirectional internal squeezing provides a practical route to approach the fundamental quantum limits set by internal dissipation in optical interferometers, without the trade-offs of major reconfigurations~\cite{ZhangPRX23GravitationalWaveDetector}, and may therefore also be valuable for other precision interferometric experiments that operate in a quantum-noise limited regime~\cite{VermeulenCQG21ExperimentObservinga}. 

\section{Acknowledgements}
The authors gratefully acknowledge the support of the United States National Science Foundation. We additionally thank Kevin Kuns for discussions and help producing a parametric amplifier in Finesse and for sharing GWINC code capable of simulating interferometers with arbitrary signal-flow diagrams. This work was produced through support from NSF grant PHY-2317110 for the Institute of Quantum Information and Matter Physics Frontier Center. LIGO operates under Cooperative Agreement No. PHY-1764464. Advanced LIGO was built under Grant No. PHY-0823459.

\nocite{apsrev42Control}
\disablelabelkeys{}
\bibliographystyle{apsrev4-2-trunc}
\renewcommand{\selectlanguage}[1]{}
\bibliography{control,zotero_mcc}

\clearpage
\renewcommand{\appendixname}{Endmatter}
\appendix

\section{GW detector implementation details}
\subsection{Quantum Radiation Pressure}
Here we derive how internal squeezing affects the quantum radiation pressure component $K(\Omega)$, defined in~\cite{DanilishinLRR19AdvancedQuantum,BuonannoPRD01QuantumNoise,KimblePRD01ConversionConventional}.
The bare optomechanical factor $K_\text{TM}$ (see Eq. 12 of \cite{McCullerPRD21LIGOsQuantum}, Sec. 5.3 of \cite{DanilishinLRR12QuantumMeasurement}) that couples the amplitude quadrature to the phase quadrature through a pair of mirrors that form an on-resonance (``undetuned'') FP cavity in a Michelson interferometer is
\begin{align}
  K_{\text{TM}}(\Omega) = -\frac{8k P_\FP}{c M_\FP \Omega^2},
\end{align}
where $P_\text{FP}$ is the power incident on the FP mirrors.  $K_{\text{TM}}$ gets enhanced through the arm and signal recycling cavities to be the externally-referred optomechanical gain~$K_{\text{ext}}$ as
\begin{align}\label{eq:KextMap}
  K_{\text{ext}}(\Omega) &= \sqrt{\frac{\Gbint}{\Gfint}}\frac{2 K_{\text{TM}}(\Omega)}{D_0A_0}
  \\
  \frac{1}{D_0A_0} &\approx \frac{|1 + g'|^2}{4}\frac{T_\FP}{T_\SE}
\end{align}
The above equation uses the DC magnitude $D_0A_0$ of the transfer function of fields generated in the arm to the readout. This factor is derived and approximated in~\cref{sup:closedform}.

For a design that includes external squeezing in combination with an optimized filter cavity (see parameters in~\cref{tab:approximation_hierarchy}) which rotates the squeezing angle to produce frequency-dependent squeezing~\cite{LIGOO4DetectorCollaborationPRX23BroadbandQuantum}, the PSD from quantum radiation pressure noise (QRPN) to first order in the losses is:
\begin{align}\label{eq:SKh_QRPN}
  S^{\mathrm{K}}_h &=
\frac{K^2_{\text{ext}}(\Omega)}{2 \Gopt L^2} \frac{\Ginj + \epsinj + 4\frac{\epsfc}{T_{\rm fc}} + {\scriptstyle (1-T_\SE)}\epsseI + \epsseIV}{D_0A_0}.
\end{align}
where we additionally approximated $\Ginj(1-\epsilon_{(\cdot)})\approx \Ginj$, which does not hold if either $\Ginj$ or the losses are large.

The external optomechanical factor $K_{\text{ext}}$ expresses the dependence of the QRPN on both the arm power and the internal squeezing gains. \cref{eq:KextMap} indicates increased QRPN due to the internal bidirectional squeezing can be compensated by changing the signal and arm mirror transmissivities.

Below, we simplify the expressions using conventional approximations valid for the standard LIGO design parameters (labelled with superscript``std'')  as well as for those for a design with bidirectional squeezing (superscript ``bd''), assuming that $T^{\text{bd}}_\FP = T^{\text{bd}}_\SE \ll 1$ and using (balanced) bidirectional squeezing with $\Gint \equiv \Gbint=1/\Gfint$.
\begin{align}
  K^{\text{std}}_{\text{ext}} &= 2K_{\text{TM}}\frac{T^{\text{std}}_\SE}{T^{\text{std}}_\FP}C^{\text{std}},
  &&&
K^{\text{bd}}_{\text{ext}} &= 2K_{\text{TM}}\Gint 
\end{align}
Where the $\Omega$ dependence is elided and $C^{\text{std}} = (1 + \sqrt{1-T^{\text{std}}_\SE})^2/4$ is an $\mathcal{O}(1)$ correction factor needed when using the advanced LIGO parameters $T^{\text{std}}_\SE=32.5\%$, $T^{\text{std}}_\FP=1.48\%$.

To keep the optomechanical factor $K_{\text{ext}}$ constant when moving to a design with BD squeezing, one can choose parameters that equate the two relations for constant $M_\FP,k$.
This means that we can approximately maintain constant radiation pressure and optical gain using
\begin{align}
\Gint P^{\text{bd}}_\FP &\approx \frac{T^{\text{std}}_\SE}{T^{\text{std}}_\FP}P^{\text{std}}_\FP C^{\text{std}}, & T^{\text{bd}}_\FP &= T^{\text{bd}}_\SE
\end{align}
We can additionally impose that the optical power at the beamsplitter $P_{\text{BS}}$ should be constant. Using that~${P_\FP = \frac{2}{T_\FP}P_\BS}$, we can keep the QRPN constant with the same $P_{\text{BS}}$ under the conditions
\begin{align}
  \frac{\Gint}{T^{\text{bd}}_\FP} &= \frac{T^{\text{std}}_\SE}{(T^{\text{std}}_\FP)^2}C^{\text{std}},
  & T^{\text{bd}}_\FP &= T^{\text{bd}}_\SE,
  & P^{\text{bd}}_\BS &= P^{\text{std}}_\BS
\end{align}

With our choice of $T^{\text{bd}}_\SE=T^{\text{bd}}_\FP=2.5\%$, we find that radiation pressure and `in-bucket' sensitivity is equal to the A+ design for $\Gint=47 = 17\,\text{dB}$.

We have confirmed using our numerical model (Finesse) that the quantum noise for out proposed design is the same as LIGO A+ below 200~Hz (see \cref{fig:LIGO_curves}), and above that frequency the sensitivity is improved. Furthermore, we find that the same design parameters for the filter cavity of the external frequency-dependent squeezer~\cite{LIGOO4DetectorCollaborationPRX23BroadbandQuantum} are optimal for both the standard and the BD squeezing design, indicating that the radiation pressure term is equalized through the above procedure, except for the increased loss from the internal squeezing contribution to $\epsseI$ and $\epsseIV$.

\subsection{Internal Squeezing Level to Saturate CWB}
The procedure above enables matching the low-frequency sensitivity of a BD internal squeezing design to the current LIGO design, but leaves some parameters free. We now proceed to justify a choice for $T^{\text{bd}}_\FP$ and its resulting $\Gint$. First, our goal is to reduce the required arm power, which can promote greater freedom in other aspects of the interferometer design, notably thermal compensation and optical coatings. Emerging coating designs can provide much lower noise, though they entail higher absorptive loss that heats mirrors.

Moreover, we aim to fully saturate the CWB for the interferometers. To do so, we need the internal squeezing level to be within about 3~dB of the $\epsse$ total SE cavity loss. With external squeezing, the light entering the interferometer is already squeezed at a level $(1-\epsinj)\Ginj + \epsinj$. To ensure the squeezing level inside the SEC is at a level limited only by the losses $\epsse$, we then need
\begin{align}
\Gfint &= \frac{\epsse}{(\Ginj + \epsinj)} = \frac{2000{\cdot}10^{-6}}{(10^{-14/10} + 0.05)} \\ &\approx \frac{1}{50}\;\;\Longleftrightarrow\;\; |10\log_{10}(\Gfint)| \approx 17\,\mathrm{dB}
\end{align}
This validates our choice of $T^{\text{bd}}_\FP=2.5\%$ to saturate the CWB from internal losses while leveraging the external squeezing system implemented as part of the A+ upgrade~\cite{LIGOO4DetectorCollaborationPRX23BroadbandQuantum}.

\vspace{-3ex}
\subsection{Squeezer Cavity}

We propose the internal squeezer cavity be similar to the existing LIGO design \cite{StefszkyJPBAMOP10InvestigationDoublyresonant,ChuaOLO11BackscatterTolerant}, with a lower cavity gain by taking the squeezer input coupler transmissivity~$\Tsqz \ge 20\%$, while requiring less than 2~W of second-harmonic pump power~\cite{OelkerOO16UltralowPhase, WadeRoSI16OptomechanicalDesign} sourced by or stabilized using light filtered through the interferometer.

The losses from the four mirrors that define the OPA cavity can be as low as 10~ppm/mirror when using small beam sizes, and the transmission loss through the OPO (i.e. the OPA plus the bow-tie cavity) can have losses~$<100$~ppm. The measured escape efficiency of LIGO squeezers in O4 \cite{LIGOO4DetectorCollaborationPRX23BroadbandQuantum}, with their configuration of $\Tsqz=7\%$, suggests all loss in these systems was due to the control field coupling mirror with transmissivity $T_{\text{M2}}=.001$, and this is consistent with ${<}500\text{ppm}$ losses within the OPO cavity.

\subsection{Transverse Mode Mismatch}

An analytic derivation of~\cref{eq:mismatch_map} using the input-output formalism accounting for both the fundamental and higher-order modes is found in~\cref{sup:closedform}; additionally, see Fig.~7 in~\cite{Kuns26SqueezedState}. The vacuum entering through the higher-order-transverse-mode (HOM) coupling, with amplitude $\sqrt{\Upsilon}$,  is suppressed by the anti-resonant cavity amplitude gain to produce an effective amplitude loss $\propto \sqrt{\Upsilon T_\SE/4}$. This factor affects the squeezing cavity in both directions, which adds coherently after the squeezing operation and just before the anti-squeezing operation. This cancels the factor of four to give an overall power loss $\epssqzMM=T_\SE\Upsilon$.

\vspace{-2ex}
\newcommand{\ashtab}[1]{}
\newcommand{\tabdef}[0]{\begin{tabular}{|lll|rr|rr|}}
\begin{table}[h!]
  \centering
  \scriptsize
    \newcommand{\tdesc}[1]{\multicolumn{1}{|l}{#1} &} 
  \newcommand{\la}[0]{\ensuremath{\leftarrow}} 
  \newcommand{\hd}[1]{} 
  \newcommand{\mc}[1]{\multicolumn{2}{c|}{#1}} 
  \tabdef
    \hline
    \textbf{Description}&\textbf{Sym.} \rule{0pt}{1.2em}& \textbf{Unit} &
    \textbf{A$^+$} & \textbf{A$^+_{\text{wb}}$} & \textbf{$\hat{\text{A}}_{17}$} & \textbf{\, $\hat{\text{A}}_{22}$\, } 
    \ashtab{& \textbf{A$^\sharp$} & \textbf{A$^\sharp_{\text{wb}}$} & \textbf{$\hat{\text{A}}^\sharp_{17}$} & \textbf{\, $\hat{\text{A}}^\sharp_{22}$\, }}\\
    \hline
    \tdesc{laser wavelength ($\mu$m)} $\lambda$                  & $\mu\text{m}$ & $1064$ &         &                         & \la \ashtab{ &       &     &                        &\la}\\
    \tdesc{signal cavity length} $L_{\text{sec}}$     & m             & $55.0$ &         &                         & \la \ashtab{ &       &     &                        &\la}\\
    \tdesc{one-way arm length } $L_{\text{arm}}$                 & m             & $3995$ &         &                         & \la \ashtab{ &       &     &                        &\la}\\
    \tdesc{arm mirror mass} $M_{\text{tm}}$                      & kg            & $40$   &         &                         & \la \ashtab{ & 105   &     &                        &\la}\\
    \hline                                                                                                                                  
    \tdesc{injected laser power} $P_{\text{in}}$                 & W             & $125$  &         & $80$                    &     \ashtab{ & 250   &     & 170                    &}\\
    \tdesc{circulating arm power} $P_{\text{arm}}$               & kW            & $750$  &         & $405$                   &     \ashtab{ & 1500  &     & 860                    &}\\
    \tdesc{circulating BS power} $P_{\text{BS}}$                 & kW            & $5.3$  &         & $5.1$                   &     \ashtab{ & 10.6  &     & 10.9                   &}\\
    \tdesc{ITM power reflectivity} $T_{\text{fp}}$               & \%            & $1.48$ &         & $2.5$                   &     \ashtab{ & 1.48  &     & 2.5                    &}\\
    \tdesc{SRM power reflectivity} $T_{\text{se}}$               & \%            & $32.5$ & $5,10$  & $2.5$                   &     \ashtab{ & 32.5  & 5   & 2.5                    &}\\
    \hline                                                                                                                                  
    \tdesc{SEC internal loss} $\epssec$                          & ppm           & $500$  &         &                         & \la \ashtab{ & $500$ &     &                        & \la}  \\
    \tdesc{external injection loss} $\epsinj$              & \%            & $5$    &         &                         & \la \ashtab{ & 4     &     &                        & \la} \\
    \tdesc{detection/readout loss } $\epsdet$                    & \%            & $10$   &         &                         & \la \ashtab{ & 2.5   &     &                        & \la} \\
    \tdesc{arm round-trip loss } $\epsilon_{\text{arm}}$         & ppm           & $75$   &         &                         & \la \ashtab{ &       &     &                        & \la} \\
    \tdesc{external squeezing level } $G_{\text{inj}}$           & dB            & $14.0$ &         &                         & \la \ashtab{ &       &     &                        & \la}\\
    \hline                                                                                                                                  
    \tdesc{internal squeezer trans.} $T_{\text{sqz}}$            &  \%           &  -\ \  &         & \mc{20\%}         \hd{\ &}    \ashtab{ & -     &     &\mc{33\%}           \hd{\ &}} \\
    \tdesc{internal squeezer loss} $\epssqz$                     &  ppm          &  -\ \  &         & \mc{100,200,400}  \hd{\ &}    \ashtab{ & -     &     &\mc{100,200,400}    \hd{\ &}} \\
    \tdesc{SE full internal loss} $\epsse$                       &  ppm          &  $500$ &         & \mc{1500-4500}    \hd{\ &}    \ashtab{ & 500   &     &\mc{1100-2900}      \hd{\ &}} \\
    \tdesc{backward-path squeezing} $G_{\text{b}}$               &  dB           & -\ \   &         & -17                     & -22 \ashtab{ & -     &     &-17                     & -22}\\
    \tdesc{forward-path squeezing} $G_{\text{f}}$                &   dB          & -\ \   &         & 17                      & 22  \ashtab{ & -     &     &17                      & 22 }\\
    \hline                                                                                                                                  
    \tdesc{filter cavity length} $L_{\text{fc}}$                 & m             & $300$  &         &                         & \la \ashtab{ &       &     &                        &\la}\\
    \tdesc{filter cavity loss} $\epsilon_{\text{fc}}$ &  ppm          & $60$   &         &                         & \la \ashtab{ & 30    &     &                        &\la}\\
    \tdesc{filter cavity detuning}                               &   Hz          & $45.8$ & 16,24   &  45.8                   & 80  \ashtab{ & 40    & 14.7&40                    &68}\\
    \tdesc{filter cavity in. trans.} $T_{\text{fc}}$     &   ppm         & $1200$ & 400,600 & 1200                    & 2000\ashtab{ & 1010  & 374 &1010                     &1700}\\
    \hline
  \end{tabular}

  \caption{Parameters used for the LIGO-like interferometer model.}
  \label{tab:LIGOUP}
\end{table}
\vspace{-2ex}

\subsection{Phase Noise and Control Sidebands}
Using squeezing with high nonlinear gain will make this interferometer susceptible to dephasing noise. Dephasing noise as low as~1~mrad has been demonstrated~\cite{OelkerOO16UltralowPhase}, but relies on carefully tuned implementation of coherent control~\cite{ChelkowskiPRA07CoherentControl,VahlbruchPRL06CoherentControl}. For internal squeezing, the coherent control and pump stability must be implemented using fields that do not require additional transmissive optics in the squeezing cavity. This will present new integration challenges, but no fundamental limitations are known or expected here.

Interferometers are also controlled using RF sidebands with a significant amount of power~\cite{AasiCQG15AdvancedLIGO}. These sidebands will likely interact with the nonlinearities of the squeezing crystal. The squeezer cavity can help suppress the entry of RF sidebands into the cavity $\propto 1/\Tsqz$, presenting a trade-off with cavity enhancement~$\propto \Tsqz$ of the losses $\epssqz$.

\vspace{-2ex}
\subsection{Thermorefractive Noise}

Modern crystal squeezers in GW interferometers utilize Periodically Poled Potassium Titanyl Phosphate (PPKTP) with $<20$~µm waist sizes. Beams passing through substrates are known to experience thermorefractive noise~\cite{BraginskyPLA00ThermorefractiveNoise,BraginskyPLA04CornerReflectors}, and this noise is enhanced for small beam sizes due to the higher energy densities they entail. Using Eq. E7 from \cite{BraginskyPLA04CornerReflectors}, we find that most nonlinear crystals have thermorefractive noise around $10^{-15}\text{m}/\sqrt{\text{Hz}}$ that is approximately flat (logarithmic) in frequency below 600~Hz when the thermal diffusion length is larger than a $20~$~µm beam size. The magnitude of this noise is proportional to the thermo-optic coefficient~$\beta=dn/dT$ of the specific crystal, through which LBO and BBO crystals have a factor of 3 and 10 lower noise ASD, respectively, than PPKTP.

Optical path length noise at the crystal in the signal extraction cavity path will only create noise sidebands through the modulation of residual carrier light from imperfect destructive interference; length noise does not imprint into the signal sidebands except through radiation pressure to first order. The conversion of motion noise to RMS noise sideband power inside the internal squeezing cavity is

\begin{align}\label{eq:thermo_limit}
P_{\rm RMS} = 2k^2P \frac{4 \Gbint T_\SE}{T_{\rm SQZ} T_\FP^2} \left(\delta l_{\FP}\right)^2
\simeq 10^{-5}\text{W},
\end{align}
for an RMS motion of $\delta l = 1\,\text{fm}$, gain $\Gbint=100$, arm power $P=750\text{kW}$, and optics with $T_{\rm sqz}=20\%$,  $T_\SE=T_\FP=2\%$.

This classical noise couples in at the same location as losses inside the squeezer. To assess to what extent this noise may limit our scheme, we compute the length noise level at which this classical noise is greater than the quantum noise introduced by the loss:
\begin{align}\label{eq:Sl_lengthnoise}
S_l \le \frac{\hbar c}{(2 - T_\SE)^2 2kP_{\rm RMS}}\left(\Gbint \epssqz + \frac{4T_{\rm sqz}}{T_{\rm fc}} \epsilon_{\rm fc}\right),
\end{align}
which is roughly $10^{-15}\text{m}/{\sqrt{\text{Hz}}}$ for the current values of squeezing loss or filter cavity loss. The factor $\Gbint$ multiplying~$\epsilon_\text{sqz}$ is included to accurately compare to \ref{eq:thermo_limit}, since the power amplification and loss are on opposite sides of the gain $\Gbint$. \cref{eq:Sl_lengthnoise} is heuristic and optimistic due to other potential sources of contrast defect light and an incomplete understanding of thermorefractive noise within an active gain medium. This noise does not appear fundamental, but will pose significant integration challenges.



\clearpage
\beginsupplement

\newcommand{\ashtab}[1]{}
\newcommand{\tabdef}[0]{\begin{tabular}{|lll|rr|rr|}}

\widetext

\tagged{standalone}{
\widetext
}
\section{Analytical model for signal-referred quantum noise in a DRFPMI}\label{sup:closedform}
Here we derive the closed-form expression for the quantum noise of a general dual-recycled Fabry-Pérot Michelson interferometer (DRFPMI) configuration, including internal and external squeezing and anti-squeezing for~\cref{eq:Sqh_DRFPMI}. These computations only incorporate dynamics of the differential mode of the interferometer, i.e. fields that are equal and opposite in phase between the arms. These dynamics can be effectively described as those of a single linear three-mirror cavity, as depicted in~\cref{fig:cavity_diagram}.

The DRFPMI system comprises three mirrors with power reflectivities $R_{(\cdot)}=1-T(\cdot)$; the SRM (signal-recycling mirror, transmissivity $T_\SE$), the ITM (input test mass, transmissivity $T_\FP$), and the ETM (end test mass, $T_\text{ITM}=0$). Parametric gain stages that apply a power gain $G_{(\cdot)}$ at the input, output, and between the SRM and ITM are included. Loss ports are included at all distinguishable parts of the system, where the loss port is modelled as a beamsplitter where the input state is transmitted with amplitude transmissivity $\sqrt{1-\epsilon_{(\cdot)}}$, and commensurately the vacuum is coupled in with an amplitude $\sqrt{\epsilon_{(\cdot)}}$. The field advances in phase for a one-way propagation between the mirrors by an amount $e^{ikc\tau_{(\cdot)}/2}$, with $\tau_\FP=2L_\FP/c,\,\tau_\SE=2L_\SE/c$.

A system of linear equations can thus be formulated that gives the field at the output in terms of the input vacuum fields and the parameters of the system. For this model, we consider the signal-referred quantum noise in the signal (phase) quadrature of the output field (\cref{eq:quantum_noise}). Our numerical codes Finesse and (customized) GWINC use exact solutions. The system of equations is approximated and solved analytically here to give a compact closed-form expression for the quantum noise.

The quantum noise is the sum of all vacuum contributions at the output: $\mathcal{S}_{a_{\rm out}}(\Omega)=\sum_l |H_l(\Omega)|^2\,\epsilon_l$, where $|H_l(\Omega)|^2$ are the magnitudes of transfer functions of vacuum from each loss port $\epsilon_l$. The noise is then referred to the signal following~\cref{eq:calib_S} to yield the quantum noise PSD. 

\tagged{standalone}{
\begin{figure}[h]
\centering
\includegraphics[width=0.7\textwidth]{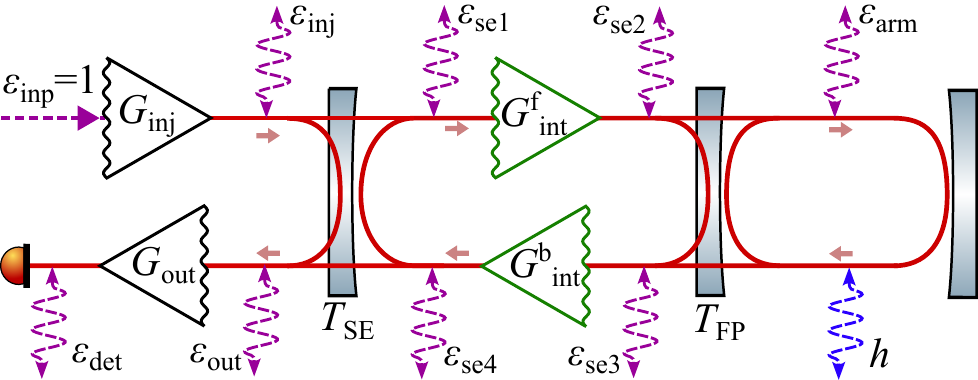}
  \caption{Unfolded signal flow diagram of a dual-recycled
    Fabry--Perot Michelson interferometer.  Upper branch: fields
    propagating to the right (forward).  Lower branch: fields
    propagating to the left (backward).  Triangles denote gain
    stages with power gain~$G_{(\cdot)}$; wavy arrows denote loss
    ports where vacuum couples in with power fraction
    $\varepsilon_{(\cdot)}$ via a beamsplitter loss model.  The
    signal~$h$ couples inside the Fabry--Perot arm cavity; the
    output is read at the detector (lower left).}
  \label{fig:cavity_diagram}
\end{figure}
}

\subsection{Transfer functions and definitions}
We start by defining the following convenience symbols for deriving the total transfer functions and losses.

\allowdisplaybreaks
\begin{align}
\ubO &\equiv (1-\epsilon_{\rm det})\,G_{\rm out}\,(1-\epsilon_{\rm out}), \\
\ubB &\equiv (1-\epsilon_{\rm se3})(1-\epsilon_{\rm se4})\,G_{\rm int}^{b}, \\
R_{\rm FP} &\equiv 1-T_{\rm FP}, \qquad
R_{\rm SE} \equiv 1-T_{\rm SE}, \\
r_{\rm FP} &\equiv \sqrt{R_{\rm FP}}, \qquad
r_{\rm SE} \equiv \sqrt{R_{\rm SE}}, \qquad
t_{\rm FP} \equiv \sqrt{T_{\rm FP}}, \qquad
t_{\rm SE} \equiv \sqrt{T_{\rm SE}}, \\
g &\equiv \sqrt{\Gfint\Gbint} \\
\alpha &\equiv \sqrt{1-\epsilon_{\rm arm}}, \\
\phi_{\FP}(\Omega) &\equiv \frac{2\Omega L_{\FP}}{c}, \\
A(\Omega) &\equiv
\left|
\frac{t_{\rm FP}}
{1-r_{\rm FP}\alpha e^{i\phi_{\FP}(\Omega)}}
\right|^{2}
=
\frac{T_{\rm FP}}
{1+R_{\rm FP}(1-\epsilon_{\rm arm})
-2r_{\rm FP}\sqrt{1-\epsilon_{\rm arm}}\cos\phi_{\FP}(\Omega)}, \\
r^{\rm eff}_\FP(\Omega)
&=
\frac{r_{\rm FP}-\sqrt{1-\epsilon_{\rm arm}}\,e^{i\phi_{\rm FP}}}
{1-r_{\rm FP}\sqrt{1-\epsilon_{\rm arm}}\,e^{i\phi_{\rm FP}}},\\
\phi_{\rm refl}^{\FP}(\Omega) &= \arg\left(r_{\FP}^{\rm eff}(\Omega)\right) \\
R_{\FP}^{\rm eff}(\Omega) &\equiv
1-\epsilon_{\rm arm}A(\Omega) \nonumber\\
&=
\frac{
R_{\rm FP}+(1-\epsilon_{\rm arm})
-2r_{\rm FP}\sqrt{1-\epsilon_{\rm arm}}\cos\phi_{\FP}(\Omega)
}{
1+R_{\rm FP}(1-\epsilon_{\rm arm})
-2r_{\rm FP}\sqrt{1-\epsilon_{\rm arm}}\cos\phi_{\FP}(\Omega)
}, \\
\Phi(\Omega) &\equiv \frac{2\Omega L_{\rm SE}}{c} + \phi_{\rm refl}^{\FP}(\Omega) + \phi_{\rm SRM}, \\
D(\Omega) &\equiv
\frac{T_{\rm SE}}
{\left|
1-r_{\rm SE}e^{i\Phi(\Omega)}
\sqrt{(1-\epsilon_{\rm se1})(1-\epsilon_{\rm se2})(1-\epsilon_{\rm se3})(1-\epsilon_{\rm se4})
\,G_{\rm int}^{f}G_{\rm int}^{b}}
\,
\frac{r_{\rm FP}-\alpha e^{i\phi_{\FP}(\Omega)}}
{1-r_{\rm FP}\alpha e^{i\phi_{\FP}(\Omega)}}
\right|^{2}}, \\
R_{\rm SE}^{\rm eff}(\Omega) &\equiv
\left|
\frac{
r_{\rm SE}
-
e^{i\Phi(\Omega)}
\sqrt{(1-\epsilon_{\rm se1})(1-\epsilon_{\rm se2})(1-\epsilon_{\rm se3})(1-\epsilon_{\rm se4})
\,G_{\rm int}^{f}G_{\rm int}^{b}}
\,
\frac{r_{\rm FP}-\alpha e^{i\phi_{\FP}(\Omega)}}
{1-r_{\rm FP}\alpha e^{i\phi_{\FP}(\Omega)}}
}{
1-r_{\rm SE}e^{i\Phi(\Omega)}
\sqrt{(1-\epsilon_{\rm se1})(1-\epsilon_{\rm se2})(1-\epsilon_{\rm se3})(1-\epsilon_{\rm se4})
\,G_{\rm int}^{f}G_{\rm int}^{b}}
\,
\frac{r_{\rm FP}-\alpha e^{i\phi_{\FP}(\Omega)}}
{1-r_{\rm FP}\alpha e^{i\phi_{\FP}(\Omega)}}
}
                               \right|^{2}.
\end{align}
Where $\phi_{\rm SRM} \in [0, \pi]$ is the SRM detuning angle. $R_{\FP}^{\rm eff}(\Omega)$ is the effective reflectivity for a field incident on the ITM, which accounts for the interference between the field that is promptly reflected off the ITM and the field that transmits through the ITM, circulates near resonance, and transmits backward through the ITM. Similarly, $R_{\rm SE}^{\rm eff}(\Omega)$ is the effective reflectivity for fields incident on the SRM which accounts for interference between promptly reflected fields and fields circulating inside both the arm and SE cavities. $A(\Omega)$ and $D(\Omega)$ are the spectral shape functions that define the frequency dependence of the transfer functions of the arm cavity and the SEC, respectively.  

We now compose the transfer functions from source to the detector for all vacua and the signal using these definitions. For the proceeding calculating, we approximate these transfer functions to first order in the losses, dropping any term $\mathcal{O}\left(\epsilon_{l}\epsilon_m\right)$ or higher. The exact transfer functions and their approximations are in~\cref{tab:transferfunctions}, which show transfer functions as a product of sub-transfer functions and a source term $\epsilon_{(\cdot)}$. We note the table includes the transfer functions for the principal external vacuum $\hat{a}_{\rm inp}$, which enters not through dissipation but through the open input port with source term $\epsilon_{\rm inp}=1$, and that of the signal with source term $h$, where $2 \Gopt L^2$ is the optical coupling transfer function of signal strain to signal sideband fields given a signal of strain amplitude $h$.
\begin{center}
\begin{table*}[h!]
\begin{tblr}{
    colspec={l|l}, 
    hlines         
}
Source & Exact $|H_l(\Omega)|^{2}\epsilon_l$ & First order in losses \\
\hline
$\hat{a}_{\rm inp}$ &
$\ubO\,R_{\rm SE}^{\rm eff}(\Omega)\,G_{\rm inj}\,(1-\epsilon_{\rm inj})$ &
$G_{\rm out}\,R_{\rm SE}^{\rm eff}(\Omega)\,G_{\rm inj}\,(1-\epsilon_{\rm inj})$ \\

$\hat{a}_{\rm inj}$ &
$\ubO\,R_{\rm SE}^{\rm eff}(\Omega)\,\epsilon_{\rm inj}$ &
$G_{\rm out}\,R_{\rm SE}^{\rm eff}(\Omega)\,\epsilon_{\rm inj}$ \\

$\hat{a}_{\rm se1}$ &
$\ubO\,D(\Omega)\,\ubB\,R_{\FP}^{\rm eff}(\Omega)\,(1-\epsilon_{\rm se2})\,G_{\rm int}^{f}\,\epsilon_{\rm se1}$ &
$G_{\rm out}\,D(\Omega)\,G_{\rm int}^{b}G_{\rm int}^{f}\,\epsilon_{\rm se1}$ \\

$\hat{a}_{\rm se2}$ &
$\ubO\,D(\Omega)\,\ubB\,R_{\FP}^{\rm eff}(\Omega)\,\epsilon_{\rm se2}$ &
$G_{\rm out}\,D(\Omega)\,G_{\rm int}^{b}\,\epsilon_{\rm se2}$ \\

$\hat{a}_{\rm arm}$ &
$\ubO\,D(\Omega)\,\ubB\,A(\Omega)\,\epsilon_{\rm arm}$ &
$G_{\rm out}\,D(\Omega)\,G_{\rm int}^{b}\,A(\Omega)\,\epsilon_{\rm arm}$ \\

$\hat{a}_{\rm se3}$ &
$\ubO\,D(\Omega)\,(1-\epsilon_{\rm se4})\,G_{\rm int}^{b}\,\epsilon_{\rm se3}$ &
$G_{\rm out}\,D(\Omega)\,G_{\rm int}^{b}\,\epsilon_{\rm se3}$ \\

$\hat{a}_{\rm se4}$ &
$\ubO\,D(\Omega)\,\epsilon_{\rm se4}$ &
$G_{\rm out}\,D(\Omega)\,\epsilon_{\rm se4}$ \\

$\hat{a}_{\rm out}$ &
$(1-\epsilon_{\rm det})\,G_{\rm out}\,\epsilon_{\rm out}$ &
$G_{\rm out}\,\epsilon_{\rm out}$ \\

$\hat{a}_{\rm det}$ &
$\epsilon_{\rm det}$ &
$\epsilon_{\rm det}$ \\

$h$ &
$\ubO\,D(\Omega)\,\ubB\,A(\Omega) (2 \Gopt L^2) $ &
$G_{\rm out}\,D(\Omega)\,G_{\rm int}^{b}\,A(\Omega) (2 \Gopt L^2) $ \\
\end{tblr}
\caption{Table with transfer functions from source to detector for all vacua and the signal, factored into terms. The second column drops explicit terms that are second order in the losses. The sources are noted as quantum operators representing vacuum bath channels that introduce quantum noise by their contributions to $\hat{a}_{\rm meas}$.}
\label{tab:transferfunctions}
\end{table*}
\end{center}

The transfer functions in the second column of~\cref{tab:transferfunctions} are linear in the losses $\epsilon_{(\cdot)}$ when we approximate each of the factors (sub-transfer functions) that compose them to zeroth order in the losses. Approximations to first order in loss for the loss-dependent factors of the total transfer functions are given in~\cref{tab:approximation_hierarchy}.
%
\begin{table}
\begin{tblr}{
    colspec={l|l|l},
    hlines
}
First order in loss
&
Zeroth order in loss
&
\makecell{Zeroth order in loss +\\single-pole/high-finesse ($\phi_\text{SRM} = \pi$)}
\\\hline

$\displaystyle
\ubO \approx
G_{\rm out}\bigl(1-\epsilon_{\rm det}-\epsilon_{\rm out}\bigr)
$
&
$\ubO \approx G_{\rm out}$
&
$\leftarrow$
\\

$\displaystyle
\ubB 
\approx
G_{\rm int}^{b}\bigl(1-\epsilon_{\rm se3}-\epsilon_{\rm se4}\bigr)
$
&
$\ubB \approx G_{\rm int}^{b}$
&
$\leftarrow$
\\

$\displaystyle
\alpha
\approx
1-\frac{\epsilon_{\rm arm}}{2}
$
&
$\alpha \approx 1$
&
$\leftarrow$
\\

$\displaystyle
\epsilon_{\rm rt}
=
\epsilon_{\rm se1}+\epsilon_{\rm se2}+A_{0}\epsilon_{\rm arm}+\epsilon_{\rm se3}+\epsilon_{\rm se4}
$
&
$\epsilon_{\rm rt}\approx 0$
&
$\leftarrow$
\\

$\displaystyle
\kappa
=
1-\frac{\epsilon_{\rm rt}}{2}
$
&
$\kappa \approx 1$
&
$\leftarrow$
\\

$\displaystyle
\rho
=
r_{\rm SE}g\,\kappa
$
&
$\rho \approx r_{\rm SE}g$
&
$\leftarrow$
\\

$\displaystyle
\beta=r_{\rm FP}\alpha
\approx
r_{\rm FP}\!\left(1-\frac{\epsilon_{\rm arm}}{2}\right)
$
&
$\beta \approx r_{\rm FP}$
&
$\leftarrow$
\\

$\displaystyle
\gamma_{\FP}
\approx
\frac{c\,(1-r_{\rm FP}\alpha)}{2\,r_{\rm FP}\alpha\,L_{\rm arm}}
$
&
$\displaystyle
\gamma_{\FP}
\approx
\frac{c\,(1-r_{\rm FP})}{2\,r_{\rm FP}\,L_{\rm arm}}
$
&
$\displaystyle
\gamma_{\FP}
\approx
\frac{T_{\FP}}{2\tau_{\FP}}
$
\\

$\displaystyle
\gamma_{{\rm SE}}
\approx
\gamma_{\FP}\left|
\frac{1-r_{\rm SE}g\,\kappa\,e^{i\phi_{\rm SRM}}}
     {1+r_{\rm SE}g\,\kappa\,e^{i\phi_{\rm SRM}}}
\right|
$
&
$\displaystyle
\gamma_{\rm SE}
\approx
\gamma_{\FP}\left|
\frac{1-r_{\rm SE}g\,e^{i\phi_{\rm SRM}}}
     {1+r_{\rm SE}g\,e^{i\phi_{\rm SRM}}}
\right|
$
&
$\displaystyle
\gamma_{\rm SE}
\approx
\frac{T_{\rm FP}}{2\,\tau_{\FP}}\frac{|1+\grtint'|}{|1-\grtint'|}
$
\\

$\displaystyle
A_{0}
\approx
\frac{T_{\rm FP}}{(1-r_{\rm FP}\alpha)^{2}}
$
&
$\displaystyle
A_{0}
\approx
\frac{T_{\rm FP}}{(1-r_{\rm FP})^{2}}
$
&
$\displaystyle
A_{0}
\approx
\frac{4}{T_{\FP}}
$
\\

$\displaystyle
A(\Omega)
\approx
\frac{T_{\rm FP}}
{\left|1-r_{\rm FP}\alpha\,e^{2i\Omega L_{\rm FP}/c}\right|^{2}}
$
&
$\displaystyle
A(\Omega)
\approx
\frac{T_{\rm FP}}
{\left|1-r_{\rm FP}e^{2i\Omega L_{\rm FP}/c}\right|^{2}}
$
&
$\displaystyle
A(\Omega)
\approx
\frac{A_{0}}{1+\Omega^{2}/\gamma_{\rm FP}^{2}}
$
\\

$\displaystyle
R_{\rm FP}^{\rm eff}(\Omega)
\approx
1-\epsilon_{\rm arm}
\frac{T_{\rm FP}}
{\left|1-r_{\rm FP}e^{2i\Omega L_{\rm FP}/c}\right|^{2}}
$
&
$R_{\rm FP}^{\rm eff}\approx 1$
&
$R_{\rm FP}^{\rm eff}\approx 1$
\\

$\displaystyle
D_{0}
\approx
\frac{T_{\rm SE}}
{\left|1-r_{\rm SE}g\,\kappa\,e^{i\phi_{\rm SRM}}\right|^{2}}
$
&
$\displaystyle
D_{0}
\approx
\frac{T_{\rm SE}}
{\left|1-r_{\rm SE}g\,e^{i\phi_{\rm SRM}}\right|^{2}}
$
&
$\leftarrow$
\\

$\displaystyle
D(\Omega)
\approx
\frac{T_{\rm SE}}
{\left|1+r_{\rm SE}e^{i\Phi(\Omega)}g\,
\frac{r_{\rm FP}-\alpha e^{2i\Omega L_{\rm FP}/c}}
     {1-r_{\rm FP}\alpha e^{2i\Omega L_{\rm FP}/c}}
\right|^{2}}
$
&
$\displaystyle
D(\Omega)
\approx
\frac{T_{\rm SE}}
{\left|1-r_{\rm SE}g\,e^{i\Phi(\Omega)}\right|^{2}}
$
&
$\displaystyle
D(\Omega)
\approx
D_{0}\,
\frac{1+\Omega^{2}/\gamma_{\rm FP}^{2}}
     {1+\Omega^{2}/\gamma_{\rm SE}^{2}}
$
\\

$\displaystyle
R_{\rm SE}^{\rm eff}(\Omega)
\approx
1-\bigl(1-g^{2}\bigr)\,D(\Omega)
$
&
$\leftarrow$
&
$\leftarrow$
\\

$A_{0}D_{0}$
&
$A_{0}D_{0}$
&
$\displaystyle
A_{0}D_{0}
\approx
\frac{4 T_\SE}{T_{\rm FP}|1+\grtint'|^2}
$
\\

$\displaystyle
D(\Omega)A(\Omega)
$
&
$D(\Omega)A(\Omega)$
&
$\displaystyle
D(\Omega)A(\Omega)
\approx
A_{0}D_{0}\,
\frac{1}{1+\Omega^{2}/\gamma_{\rm SE}^{2}}
$
\\
\end{tblr}
\caption{Approximations for the transfer function factors including the spectral factors. The first column keeps terms to first order in loss, the second only keeps terms zeroth order in loss such that the total noise $S^q_h$ is first order in loss, and the third column additionally applies the single-pole and high-finesse approximations where applicable.}
\label{tab:approximation_hierarchy}
\end{table}

Note that the DRFPMI configuration can also be depicted by the signal flow diagrams~\cref{fig:cavity_diagram} discussed in the following section, which express the system of equations for the propagation of optical fields through the coupled cavities. Nodes represent variables and edges represent coupling coefficients (i.e. a matrix). The graph connectivity represents the dependencies between variables. Forward and backward propagating fields are depicted along upper and lower branches, respectively. Mirror reflections couple the upper and lower branches. The linear system of equations can be solved through Gaussian elimination that removes loops. The ordering of loop eliminations emphasizes different cavity physics. Here, we reduce the arm loop first, giving a transfer function $A(\Omega)$ from the arm to the SEC. Then reducing the SEC produces $D(\Omega)$. This reduction ordering allows the product $A(\Omega)D(\Omega)$ to be the transfer function of signals produced in the arm to the readout port.

\subsection{Noise terms by origin}
We now proceed to compose parts of the total noise $\mathcal{S}_{a_{\rm out}}(\Omega)=\sum_l |H_l(\Omega)|^2\,\epsilon_l$, grouping noise terms by origin on the injection path $S_{\rm inj}$ (left of the SRM on the upper branch), between the SRM and the ETM $S_{\rm cav}$, and on the readout path $S_{\rm ro}$ (left of the SRM on the lower branch):
\begin{align}
S_{\rm inj}
&=
\ubO\,R_{\rm SE}^{\rm eff}
\left[(1-\epsilon_{\rm inj})G_{\rm inj}+\epsilon_{\rm inj}\right],
\\
S_{\rm cav}
&=
\ubO\,D\Bigl[\ubB\bigl(R_{\rm FP}^{\rm eff}\,G_f\,\epsilon_{\rm se1}+\epsilon_{\rm se2}+A\,\epsilon_{\rm arm}\bigr)+G_b\,\epsilon_{\rm se3}+\epsilon_{\rm se4}\Bigr],
\\
S_{\rm ro}
&=
(1-\epsilon_{\rm det})G_{\rm out}\,\epsilon_{\rm out}+\epsilon_{\rm det}.
\end{align}
We plug in the transfer functions to find simple closed form expressions, starting with the injection term:
\begin{align}
S_{\rm inj}
&=
\ubO\Bigl[1-(1-g^2)D\Bigr]
\Bigl[(1-\epsilon_{\rm inj})G_{\rm inj}+\epsilon_{\rm inj}\Bigr],
\\
&=
\ubO\Bigl((1-\epsilon_{\rm inj})G_{\rm inj}+\epsilon_{\rm inj}
-D(1-g^2)\bigl[(1-\epsilon_{\rm inj})G_{\rm inj}+\epsilon_{\rm inj}\bigr]\Bigr).
\end{align}
For the cavity term,
\begin{align}
S_{\rm cav}
&=
\ubO D \ubB\Bigl[R_{\rm FP}^{\rm eff}\,G_f\,\epsilon_{\rm se1}+\epsilon_{\rm se2}+A\,\epsilon_{\rm arm}+G_b\,\epsilon_{\rm se3}+\epsilon_{\rm se4}/B\Bigr],\\
&\approx
\ubO D \ubB A\left[\frac{
G_f\,\epsilon_{\rm se1}+\epsilon_{\rm se2}}{A}
+\epsilon_{\rm arm}
+\frac{G_b\,\epsilon_{\rm se3}+\epsilon_{\rm se4}}{A \ubB}
\right],
\end{align}
where we dropped first-order-in-loss terms in $R_{\rm FP}^{\rm eff}$ to keep the full expression first order in loss. For the readout term,
\begin{align}
S_{\rm ro}
&=
(1-\epsilon_{\rm det})G_{\rm out}\,\epsilon_{\rm out}+\epsilon_{\rm det},
\\
\end{align}
We refer each term to the signal by dividing by the signal transfer function. The signal-referred spectrum is thus
\begin{align}
S^q_h
=\frac{1}{2}
\left[
\frac{S_{\rm inj}}{|H_h|^2}
+\frac{S_{\rm cav}}{|H_h|^2}
+\frac{S_{\rm ro}}{|H_h|^2}
\right],
\qquad
|H_h|^2=\ubO D \ubB A (2 \Gopt L^2).
\end{align}
Therefore,
\begin{align}
\frac{S_{\rm inj}}{|H_h|^2}
&= \frac{1}{2 \Gopt L^2}\Biggl[
\frac{G_{\rm inj}-(G_{\rm inj}-1)\epsilon_{\rm inj}}{\ubB D A}
-\frac{(1-g^2)\bigl[(1-\epsilon_{\rm inj})G_{\rm inj}+\epsilon_{\rm inj}\bigr]}{\ubB A}\Biggr],
\\
\frac{S_{\rm cav}}{|H_h|^2}
&=\frac{1}{2 \Gopt L^2}\Biggl[
\frac{G_f\,\epsilon_{\rm se1}+\epsilon_{\rm se2}}{A}
+\epsilon_{\rm arm}
+\frac{G_b\,\epsilon_{\rm se3}+\epsilon_{\rm se4}}{A \ubB}\Biggr],
\\
\frac{S_{\rm ro}}{|H_h|^2}
&=\frac{1}{2 \Gopt L^2}\Biggl[
\frac{\epsilon_{\rm out}}{\ubB D A}
+\frac{\epsilon_{\rm det}}{\ubB D A \ubO}\Biggr].
\end{align}
We now group terms by their frequency dependence (through $A(\Omega), D(\Omega)$) and plug in transfer functions,
\begin{align}
\Sigma_{\rm ext}
&=
\frac{1}{\Gbint}\left[G_{\rm inj}-(G_{\rm inj}-1)\epsilon_{\rm inj}+\epsilon_{\rm out}+\epsilon_{\rm det}/\Gout\right],
\\
\Sigma_{\rm int}
&=
\Gfint\epsseI
+\epsilon_{\rm se2} + \epsseIII
+\frac{1}{\Gbint}\left[\epsseIV-\left(1-g^2\right)\Bigl(G_{\rm inj}-\epsilon_{\rm inj}(G_{\rm inj}-1)\Bigr)\right].
\end{align}
Which gives
\begin{align}
 S^q_h(\Omega)
  = \frac{1}{2 \Gopt L^2}\left[\frac{1}{D(\Omega)A(\Omega)}
\Sigma_{\mathrm{ext}}
  + \;\frac{1}{A(\Omega)}
\Sigma_{\mathrm{int}}
  + \epsarm
  \right],
\end{align}

\subsection{Spectral shape terms $A(\Omega)$ and $D(\Omega)$}
The frequency dependence of the quantum noise is fully determined by the cavity power transfer functions $A(\Omega)$ and $D(\Omega)$, which describe the FP arm cavity and SEC respectively. To find a compact closed form expression for the total noise for frequencies far below the free spectral range of the arm cavity ($c/(2L_\FP)$), we approximate these functions as Lorentzians, or equivalently we replace them with their single-pole Padé approximant. Starting with the exact form of $A(\Omega)$, 
\begin{align}
A(\Omega)
&=
\left|\frac{t_{\rm FP}}{1-r_{\rm FP}\alpha e^{i\phi_{\FP}(\Omega)}}\right|^2,
\qquad
\phi_{\FP}(\Omega)=\frac{2\Omega L_{\FP}}{c}.
\end{align}
For frequencies $\Omega\ll1/\tau_{\text{arm}}=c/(2L_\text{arm})$, $\phi=\Omega\ll1$,
\begin{align}
A(\Omega)
&\approx
\left|\frac{t_{\rm FP}}{1-r_{\rm FP}\alpha\left(1+i\phi_{\FP}\right)}\right|^2
\\
&=
\left|\frac{t_{\rm FP}}{(1-r_{\rm FP}\alpha)-i\,r_{\rm FP}\alpha\,\phi_{\FP}}\right|^2
\\
&=
\left|\frac{t_{\rm FP}}{1-r_{\rm FP}\alpha}\right|^2
\frac{1}{\left|1-i\,\dfrac{r_{\rm FP}\alpha\,\phi_{\FP}}{1-r_{\rm FP}\alpha}\right|^2}
\\
&=
\left|\frac{t_{\rm FP}}{1-r_{\rm FP}\alpha}\right|^2
\frac{1}{1+\Omega^2/\gamma_{\FP}^2},
\end{align}
where we have written the function as the product of a DC gain $A_0$ and a Lorentzian function with FWHM $\gamma_\FP$, where 
\begin{align}
A_0\equiv \frac{T_{\rm FP}}{\left|1-r_{\rm FP}\alpha\right|^2}, \qquad \gamma_{\FP}=\frac{1-r_{\rm FP}\alpha}{2r_{\rm FP}\alpha\,\tau_{\FP}}.
\end{align}

For the signal-extraction cavity response we have
\begin{align}
D(\Omega)
&=
\left|\frac{T_{\rm SE}}{1-r_{\rm SE}\,\rho\,e^{i\Phi(\Omega)}}\right|^2,
\end{align}
and we define
\begin{align}
\beta&=r_{\rm FP}\,\alpha, \qquad \rho = r_\SE \kappa g,\\
\kappa &= 1 - \frac{\epsilon_\text{rt}}{2} = 1- \frac{1}{2}\left( \epsseI + \epsseII + A_0\epsarm + \epsseIII + \epsseIV \right).
\end{align}
We start by approximating $2\Omega L_\SE/c \ll \phi^{\FP}_{\rm eff}\approx 2 \arctan(\Omega/\gamma_{\FP})$, so that
\begin{align}
e^{i\Phi}
&\approx
 e^{i\left(\phi_{\rm SRM}+2\arctan(\Omega/\gamma_{\FP})\right)}
=
e^{i\phi_{\rm SRM}}\frac{\gamma_{\FP}+i\Omega}{\gamma_{\FP}-i\Omega}.
\end{align}
Thus
\begin{align}
D(\Omega)
&=
\frac{T_{\rm SE}(\gamma_{\FP}^2+\Omega^2)}{
\left|\gamma_{\FP}(1-\rho e^{i\phi_{\rm SRM}})-i\Omega(1+\rho e^{i\phi_{\rm SRM}})\right|^2
}.
\end{align}

As the closed form expression depends on the product $A(\Omega)D(\Omega)$, we seek to write this product as a single Lorentzian function. We have
\begin{align}
A(\Omega)
&=
A_0\frac{\gamma_{\FP}^2}{\gamma_{\FP}^2+\Omega^2},
\qquad
A_0=\frac{T_{\rm FP}}{(1-\beta)^2},
\end{align}
therefore,
\begin{align}
D(\Omega)A(\Omega)
&=
\frac{T_{\rm SE}T_{\rm FP}\,\gamma_{\FP}^2}{(1-\beta)^2}
\frac{1}{
\left|\gamma_{\FP}(1-\rho e^{i\phi_{\rm SRM}})-i\Omega(1+\rho e^{i\phi_{\rm SRM}})\right|^2
}.
\end{align}
We define for convenience
\begin{align}
z_0&=1-\rho e^{i\phi_{\rm SRM}},
\qquad
z_1=1+\rho e^{i\phi_{\rm SRM}},
\end{align}
so that
\begin{align}
D(\Omega)A(\Omega)
&=
A_0T_{\rm SE}\gamma_{\FP}^2
\frac{1}{\gamma_{\FP}^2|z_0|^2+\Omega^2|z_1|^2}
\\
&=
A_0T_{\rm SE}\gamma_{\FP}^2
\frac{1}{|z_0|^2\left(\gamma_{\FP}^2+\Omega^2\,\left|\frac{z_1}{z_0}\right|^2\right)}
\\
&=
A_0T_{\rm SE}\frac{\gamma_{\FP}^2}{|z_0|^2\gamma_{\rm SE}^2}
\frac{1}{1+\Omega^2/\gamma_{\rm SE}^2}
\\
&=
A_0D_0\frac{1}{1+\Omega^2/\gamma_{\rm SE}^2}.
\end{align}
Here
\begin{align}
D_0&=\frac{1}{|1-r_\SE g \kappa e^{i\phi_{\rm SRM}}|^2},
\end{align}
and
\begin{align}
\gamma_{\rm SE}
&=
\gamma_{\FP}\left|\frac{1-r_\SE g \kappa e^{i\phi_{\rm SRM}}}{1+r_\SE g \kappa e^{i\phi_{\rm SRM}}}\right|.
\end{align}

\subsection{Final result}
We summarise our final compact closed-form expression for the signal-referred quantum noise in a general DRFPMI configuration that employs squeezing and amplification. The quantum noise can in general be written as
\begin{align}
 S^q_h(\Omega)
  = \frac{1}{2 \Gopt L^2}\left[\left[A(\Omega)\right]^{-1}
    \Sigma_{\mathrm{int}}
  + \epsarm
  +
  \left[D(\Omega)A(\Omega)\right]^{-1}
  \Sigma_{\mathrm{ext}}
  \right],
\end{align}
where $\Sigma_{\mathrm{ext}}, \Sigma_{\mathrm{int}}$ are as defined above, which parametrise loss contributions originating internal and external to the cavities that comprise the DRFPMI. The frequency dependence of the noise is given by two functions $A(\Omega),D(\Omega)$, which are the transfer functions of the FP arm cavity and the SEC respectively. For frequencies far below the FSR of the arm cavity, we can write these functions and their products as Lorentzians with FWHM $\gamma_{(\cdot)}$; i.e. as a Padé approximant with a single pole $\gamma_{(\cdot)}$ and DC gains $A_0,D_0$, such that
\begin{align}\label{eq:sup_Sqhsp}
 S^q_h(\Omega)
  = \frac{1}{2 \Gopt L^2}\left[\frac{1}{A_0}
    \Bigl(1+\frac{\Omega^2}{\gamma_{\FP}^2}\Bigr)\Sigma_{\mathrm{int}}
  + \epsarm
  +\frac{1}{D_0A_0}
    \Bigl(1+\frac{\Omega^2}{\gamma_\text{SE}^2}\Bigr)\Sigma_{\mathrm{ext}}
  \right],
\end{align}
\subsubsection{Spectral shapes to first order in loss}
We can approximate $A(\Omega), A(\Omega)D(\Omega)$ to first order in the losses by approximating the DC gains and poles as
\begin{align}
A_0 D_0
=
\frac{T_{\rm FP}T_{\rm SE}}
{(1-r_{\rm FP})^2\,\left|1-r_{\rm SE} g e^{i\phi_{\rm SRM}}\right|^2}, \qquad A_0=\frac{T_{\FP}}{(1 - r_\FP)^2} 
\end{align}
\begin{align}
    \gamma_{\FP}=\frac{1-r_{\rm FP}}{r_{\rm FP}\,\tau_{\FP}},
\qquad
\gamma_{\rm SE}
=
\gamma_{\FP}
\left|
\frac{1-r_{\rm SE}g\,e^{i\phi_{\rm SRM}}}
     {1+r_{\rm SE}g\,e^{i\phi_{\rm SRM}}}
\right|
\end{align}
\subsubsection{Spectral shapes for $T_\FP\ll1$, $T_\SE\ll1$ (high-finesse), $\epsilon_{{(\cdot)}}\rightarrow0$ (zero loss)}
We further approximate $A(\Omega),D(\Omega)$ to zeroth order in the losses so that the full closed form expression $S^q_h$ is first order in loss. We now also assume $T_\FP\ll1$, $T_\SE\ll1$. We find for $A(\Omega)$:
\begin{align}
\gamma_{\FP} \approx \frac{T_\FP}{2\tau_{\FP}}, \quad A_0 \approx \frac{4}{T_\FP},
\end{align}
So
\begin{align}
    1/A(\Omega) = \left(\frac{T_\FP}{4} + \frac{\Omega^2\tau_\text{arm}^2}{T_\FP} \right)
\end{align}
And for $A(\Omega)D(\Omega)$, for a signal-extraction configuration with $\phi_{\rm SRM}=\pi$:
\begin{align}
 \gamma_{\rm SE}
    \approx \frac{T_{\rm FP}}{2\,\tau_{\FP}}\frac{|1+\grtint'|}{|1-\grtint'|},
    \qquad     
D_0 A_0 \approx
    \frac{4 T_\SE}{T_{\rm FP}|1+\grtint'|^2}, 
\end{align}
where we incorporate a corrected form of the SEC round-trip: $\grtint'=\grtint\sqrt{1-T_\SE}$, to get
\begin{align}
  \frac{1}{D(\Omega)A(\Omega)} \approx   \frac{T_{\rm FP}\,|1+\grtint'|^2}{4T_\SE} +  \frac{\Omega^2\tau_{\text{arm}}^2|1-\grtint'|^2}{T_\FP T_\SE}.
\end{align}
Plugging the expressions in this section into~\cref{eq:sup_Sqhsp}, we obtain~\cref{eq:Sqh_DRFPMI} in the main text, which is accurate to the numerical models to within $\approx 10\%$ up to frequencies 10~kHz for the current LIGO (A+) parameters.

\clearpage

\tagged{standalone}{
\widetext
}
\section{Comparisons to Alternate Schemes}\label{sup:AlternateSchemes}
Here we apply our general DRFPMI quantum noise model to various other squeezing and amplification schemes that have been previously proposed to evaluate their performance in terms of the achievable quantum-shot-noise-limited sensitivity. We compare these to our bidirectional internal squeezing scheme, and find that only this scheme attains the most stringent Callen-Welton bounds on sensitivity from optical loss. Different squeezing schemes correspond to choosing different values for the parametric gains $G_{(.)}$, which determines the spectral contributions of loss ports to the signal-referred quantum noise. In addition, $\Gbint,\Gfint$ affect the SEC transfer function $D(\Omega)$ itself, changing the frequency dependence of the signal-referred quantum noise. We plot the quantum shot noise that can be obtained in optimal implementations of the different schemes in~\cref{fig:alt_schemes}.  

\subsection{External squeezing}
GW detectors currently use a scheme where squeezing is injected at a level limited by injection loss, and no internal or output squeezing is employed; this corresponds to setting $\Gout=\Gfint=\Gbint=1$, and $\Ginj\lesssim\epsinj/(1-\epsinj)$. An optimal internal squeezing scheme corresponds to $\Gout=\Gfint=\Gbint=1$, and $\Ginj\to 0$. Plugging this into~\cref{eq:aggregate_losses} gives
\begin{align}
\Sigma_{\rm ext}
&=
\epsinj+\epsout+\epsdet,
\\
\Sigma_{\rm int}
&=
\epsseI+\epsseII+\epsseIII+\epsseIV,
\end{align}
Such that the total quantum noise for this scheme is

\begin{align}
S_h^q(\Omega)
&=
\frac{1}{2\Gopt L^2}
\Biggl[
\left[A(\Omega)\right]^{-1}\left(
\epsseI+\epsseII+\epsseIII+\epsseIV
\right)
+\epsarm
+
\left[D(\Omega)A(\Omega)\right]^{-1}
\left(
\epsinj+\epsout+\epsdet
\right)
\Biggr].
\end{align}

\subsection{Symmetric internal squeezing (`Quantum Expander')}
For a scheme with only symmetrical internal squeezing, i.e. where $\Gout=\Ginj=1$, and $\Gfint=\Gbint=\Gint<1$, 
\begin{align}
\Sigma_{\rm ext}
&=
\frac{1+\epsout+\epsdet}{\Gint},
\\
\Sigma_{\rm int}
&=
\Gint\,\epsseI
+\epsseII
+\epsseIII
+\frac{\epsseIV-\left(1-\Gint^2\right)}{\Gint},
\end{align}
We can then choose $\Gint$ to optimise the sensitivity. This scheme has been proposed by Korobko et al., and dubbed a `Quantum Expander'~\cite{KorobkoLSA19QuantumExpander}. By setting $\Gint = \sqrt{1-T_\SE}$, we achieve minimal noise at DC while maximising the bandwidth through $D(\Omega)$. Specifically, plugging $\Gint = \sqrt{1-T_\SE}$ into $\grtint'$, the coupled cavity transfer function becomes 
\begin{align}
A(\Omega)D(\Omega)
&\approx
\frac{4T_{\SE}}{T_{\FP}(2-T_{\SE})^2}
\frac{1}{1+\Omega^2/\gamma_{\SE}^2},
\quad \text{with} \quad
\gamma_{\SE}
\approx
\frac{T_{\FP}}{2\tau_{\FP}}\frac{2-T_{\SE}}{T_{\SE}},
\end{align}
in the single-pole, high-finesse approximation (See~\cref{sup:closedform}). The signal-referred quantum noise is thus:
\begin{align}
S_h^q(\Omega)
&=
\frac{1}{2\,\Gopt L^2}
\left[
\epsarm
+
\frac{
(1-T_{\SE})\epsseI
+\sqrt{1-T_{\SE}}\,(\epsseII+\epsseIII)
+\epsseIV-T_{\SE}
}{
\sqrt{1-T_{\SE}}\,A(\Omega)
}
\right.
\\
&\qquad\left.
+
\frac{1+\epsout+\epsdet}{\sqrt{1-T_{\SE}}}
\left(
\frac{T_{\FP}(2-T_{\SE})^2}{4T_{\SE}}
+
\frac{\Omega^2\tau_{\FP}^2\,T_{\SE}}{T_{\FP}}
\right)
\right].
\end{align}

\subsection{Three-stage squeezing and amplification scheme}
It was initially proposed by Caves~\cite{CavesPRD81QuantummechanicalNoise} that given some combination of losses (injection, internal, output) an IFO configuration that employs injected squeezing, internal squeezing, and output amplification would yield the best sensitivity with any combination of losses, i.e. it would saturate the CWB. As shown by Korobko et al.~\cite{KorobkoPRA23FundamentalSensitivity}, this is true for a dual-recycled Michelson interferometer (i.e. without FP arm cavities); however, as follows from~\cref{eq:Sqh_DRFPMI}, such a scheme employed in a DRFPMI does not saturate the CWB from internal loss for that system. Setting the parametric gains to the optimal parameters $\Gout \to \infty,\, \Ginj \to 0,\, \Gfint = \Gbint = \Gint = \sqrt{1-T_{\SE}}$, the loss aggregates are
\begin{align}
\Sigma_{\rm ext}
&=
\frac{\epsinj+\epsout}{\sqrt{1-T_{\SE}}},\\
\qquad
\Sigma_{\rm int}
&=
\sqrt{1-T_{\SE}}\,\epsseI
+\epsseII
+\epsseIII
+\frac{\epsseIV-T_{\SE}\epsinj}{\sqrt{1-T_{\SE}}}.
\end{align}
Which gives
\begin{align}
S_h^q(\Omega)
&=
\frac{1}{2\,\Gopt L^2}
\left[
\epsarm
+
\frac{
(1-T_{\SE})\epsseI
+\sqrt{1-T_{\SE}}\,(\epsseII+\epsseIII)
+\epsseIV-T_{\SE}\epsinj
}{
\sqrt{1-T_{\SE}}\,A(\Omega)
}
\right.
\\
&\qquad\left.
+
\frac{\epsinj+\epsout}{\sqrt{1-T_{\SE}}}
\left(
\frac{T_{\FP}(2-T_{\SE})^2}{4T_{\SE}}
+
\frac{\Omega^2\tau_{\text{arm}}^2\,T_{\SE}}{T_{\FP}}
\right)
\right].
\end{align}
which exceeds the CWB from combined internal and external losses, and saturates the CWB from internal losses alone at high frequency, but does not saturate this ultimate bound at DC.

\subsection{Bidirectional Internal Squeezing}
Our proposed bidirectional internal squeezing scheme prescribes
$\Ginj = 1,\; \Gout = 1,\; \Gbint = 1/\Gfint \to \infty,\; \grtint^2 = \Gfint \Gbint = 1$
which gives
\begin{align}
\Sigma_{\rm ext}
&=
0,
\\
\Sigma_{\rm int}
&=
\epsseII+\epsseIII,
\end{align}
and thus the scheme saturates the CWB bound from internal losses, taking $\epssec \equiv \epsseII+\epsseIII$, we have $S^q_h = \mathcal{S}_h^{\text{CWB,int}}$.

\begin{figure*}[hbt]
    \centering
    \includegraphics[width=0.8\linewidth]{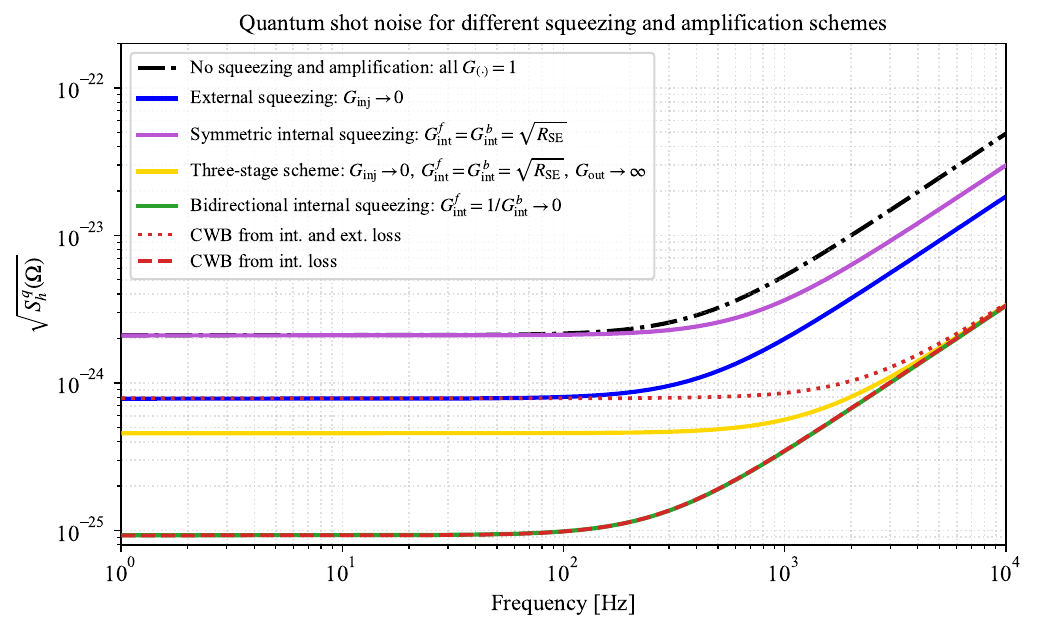}
    \caption{The quantum shot noise for a LIGO-like DRFPMI (parameters are in \cref{tab:LIGOUP}) configured with different squeezing schemes, as described above. The Callen-Welton bounds on sensitivity from optical loss are plotted for comparison.}
    \label{fig:alt_schemes}
\end{figure*}

\clearpage
\section{Derivation of Mode Healing}\label{sup:ModeHealing}

The mode healing property of bidirectional internal squeezing, whereby the introduction of vacuum states through couplings to higher-order transverse modes is suppressed, is a major advantage of the scheme. These vacua are introduced through optical dissipation from both static and thermally-induced wavefront distortion. External parametric amplification immediately after the signal-extraction cavity mirror would not have this mode healing property, and thus external amplification can be expected to be limited by mismatch loss, $\epsdet > \Upsilon$.

The derivation of mode healing on this loss mechanism requires significant linear algebra, which we depict using a signal flow diagram of~\cref{fig:signalflowgraph}. The left half of this diagram shows the system of equations for the fundamental fundamental-mode while the right shows the higher-order mode. When the mismatch parameter $\Upsilon\rightarrow 0$ then the right-side of the diagram can be ignored. In this case, the diagram is also useful for setting up and computing the transfer functions given in \cref{sup:closedform} and reducing the internal squeezing cavity to an equivalent collection of gain and loss elements in \cref{sec:intsqz_reductions}.

\begin{figure}[h]
\centering
\includegraphics[width=1.0\linewidth,page=2]{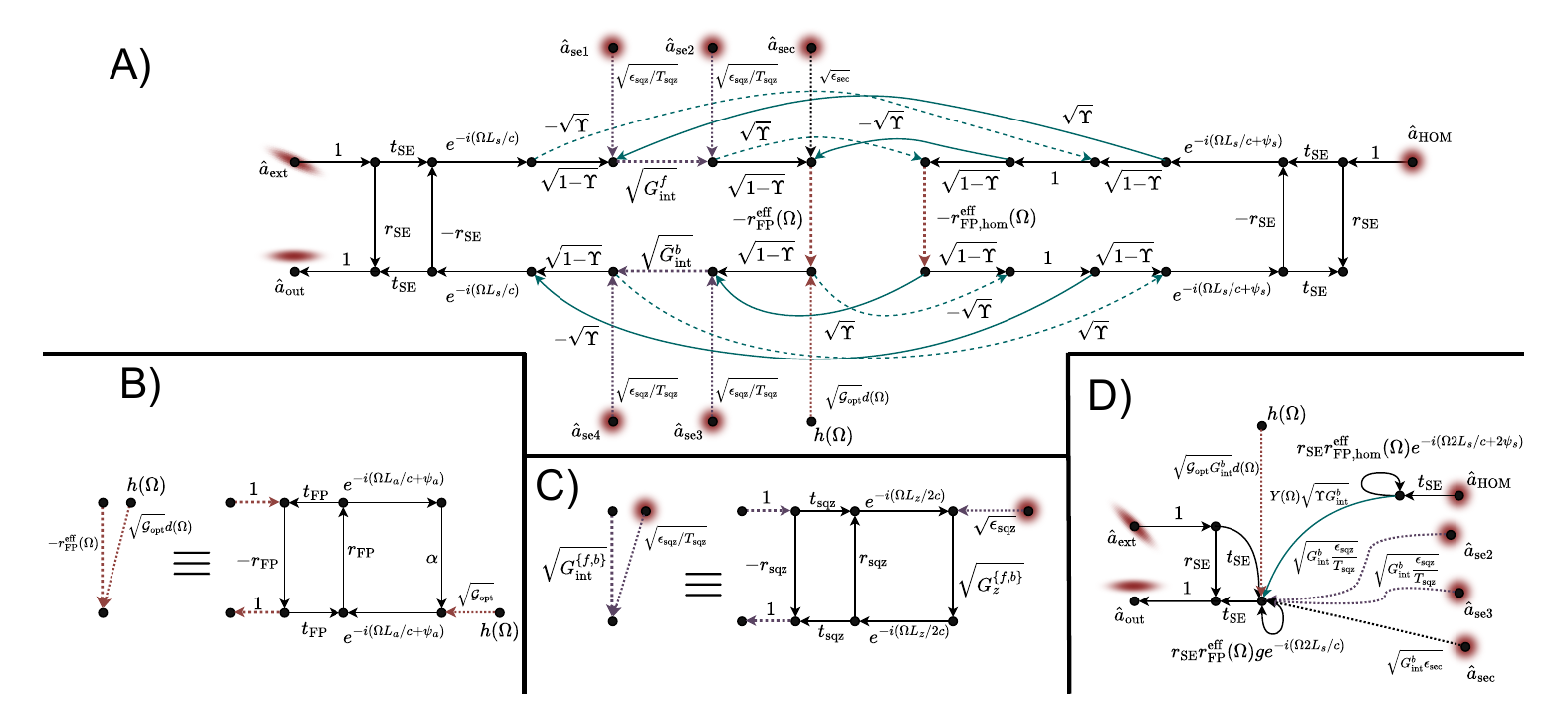}
  \caption{{\bf A)} Partially condensed signal flow diagram of a dual-recycled
    Fabry--Perot Michelson interferometer. Nodes indicate variables, $\vec{V}$, edges coefficients, $\mathbf{M}$ and the graph then represents the equation $\vec{V}=\mathbf{M}\vec{V}$. Dotted lines represent solved edges from subgraphs of the arm cavity (B) and for the internal squeezing cavity (C). The external nodes without dependencies are source terms. The source terms with fuzzy circles are from couplings to dissipation/loss that couple in vacuum states (given as operators here in a Heisenberg picture). The source term with an ellipse is from the external frequency-dependent squeezed state injection system. The source term $h(\Omega)$ represents signal imparting a phase modulation on the carrier light. The nodes can be removed through Gaussian elimination (GE), computing only the source to sink frequency-dependent scattering matrix. Higher order mode couplings are given by green lines. A perturbative solution can be found by dropping the dashed green terms. {\bf B} Subgraph of the arm Fabry-Perot cavity. Reducing this graph through Gaussian elimination gives the terms on the left. {\bf C)} Subgraph for the internal parametric amplifier cavity, discussed in following sections. {\bf D)} Reduction of graph A) into the fundamental and higher order mode cavities following the perturbative elimination of dashed green. Subdominant loss terms are omitted for space. The loss term of the HOM shows the ``off-diagonal'' coupling $Y(\Omega)\sqrt{\Upsilon G^b_{\rm int}}$ as well as the HOM loop term. Popping that loop using GE gives a factor  $D^{\rm HOM}(\Omega) \sim T_{\rm SE}/4$ to the off-diagonal coupling. Together the factors lead to the scaling \cref{eq:mismatch_map}.}
  \label{fig:signalflowgraph}
\end{figure}

The diagram A) of \cref{fig:signalflowgraph} already utilizes a partial reduction of B) and C) diagrams to form the dashed lines. First, we will analyze the reduction of B) to provide an example. The source and sink nodes with direct unit ($1$) couplings indicate a direct reflection and transmission with a cavity feedback loop. Removing all nodes of the loop 
\begin{align}
  -r_{\rm FP}^{\rm eff, gen}(\Omega) &= -r_{\rm FP} + \frac{t_{\rm FP}^2}{1 - \alpha e^{-2i\Omega L_a/c - 2i\psi_a}}
  &&&
R_{\rm FP}^{\rm eff, gen}(\Omega) &\equiv \left|r_{\rm FP}^{\rm eff, gen}(\Omega)\right|^2
      \\
  a^{\rm gen}(\Omega) &= \sqrt{\Gopt}\frac{t_{\rm FP}e^{-i\Omega L_a/c+i\psi_a}}{1 - \alpha e^{-2i\Omega L_a/c+i2\psi_a}}
  &&&
A^{\rm gen}(\Omega) &\equiv \left|a^{\rm gen}(\Omega)\right|^2
\end{align}

In the above, the $\psi_a$ term is a new free variable to express the Gouy phase of a TEM mode. For the fundamental mode, $\psi_a=0$, and $r_{\rm FP}^{\rm eff, gen} = r_{\rm FP}^{\rm eff}$ (the effective arm reflectivity). Similarly, the superscripts ``gen'' can be removed for the $a(\Omega)$ transfer functions by setting $\psi_a=0$. The arm Gouy phase in LIGO for the first HOM is about $\psi_a=48^\circ=0.83$~rad. When using this value, the gen->HOM transfer functions are computed. This Gouy phase causes the reflectivity of higher order modes from the arm cavity to be constant: $r_{\rm FP}^{\rm eff, HOM}(\Omega) \approx 1$.

The above two transfer functions can be computed from the graph by setting up the equivalent system of equations and solving them, to find the matrix directly from source to sink nodes. That matrix computed from B) is then substituted as the dashed red and purple edges of A) in the figure. In the following sections, this procedure will be performed on the C) graph to incorporate the internal squeezing cavity as an equivalent system of gains and beamsplitter losses.

Now we return to the A) and D) signal flow graphs to derive the mode healing effect. We first eliminate the dashed green lines for the coupling of the fundamental into the higher-order-mode as a perturbative approximation. We first reduce A) into the graph D) by using Gaussian Elimination (GE). At this stage, no loops are reduced so we are just propagating coupling factors. This removes most nodes and leaves two independent loops. There would be three coupled loops if the dashed lines were not removed.

GE on the left loop along with one of the transmission coefficients produces the SE cavity transfer function, $a(\Omega)$. GE on the right loop creates the arm transfer function of the HOMs:
\begin{align}
  d^{\rm HOM}(\Omega) &= \frac{t_{\rm SE}}{1 - r_{\rm SE} r^{\rm eff, HOM}_{\rm FP}(\Omega)e^{-2i\Omega L_{\rm SE}/c +2i\psi_s}} \approx \frac{t_{\rm SE}}{2}
              &&&
D^{\rm HOM}(\Omega) &\equiv \left|d^{\rm HOM}(\Omega)\right|^2
\end{align}

  The factor of $Y(\Omega)$ on the HOM coupling comes from the interference of the first and second passes of the cavity and the basis change parameterized by $\Upsilon$ in-to and out-of the squeezer cavity. It can be bounded $|Y(\Omega)| \le 2$. The full expression is 
\begin{align}
  Y(\Omega) &= -\left(1+r_{\rm FP}^{\rm eff, HOM}\right)\left(1 - \frac{1}{\sqrt{G^{\text{b}}_{\text{int}}}}\right)
\end{align}
  Where the first term represents the sign change in the coupling to/from the arm cavity and the second term are the coupling factors before/after the squeezer cavity from 4 total lines collected into the $Y$ factor.

  The total coupling from the HOM vacuum to the output is then
\begin{align}
  H^{\text{HOM}} = A(\Omega)\sqrt{G^{\text{b}}_{\text{int}}} \left|Y(\Omega)\right|^2 D^{\rm HOM}(\Omega) \Upsilon \approx A(\Omega)\sqrt{G^{\text{b}}_{\text{int}}} T_{\rm SE}\Upsilon
\end{align}
  Leading to the relation \cref{eq:mismatch_map} of the main text. This derivation is similar to the internal mismatch loss computations in \cite{Kuns26SqueezedState}.

\section{Conversion of round-trip crystal squeezing factor and intra-squeezer loss to effective squeezing factors and loss}
  \label{sec:intsqz_reductions}

  This section analyzes how to translate the travelling-wave (bow tie) cavity-enhanced squeezer into an equivalent free-space parametric amplifier that has loss before and after the amplification. The main text does not require the translation of the squeezing level, but does require computing the effect of the cavity on the losses added by the internal cavity squeezer. These translations are useful for implementations and for accurate simulations, where the effective parametric gain of the cavity must be computed using the gain at the squeezer crystal and the cavity parameters.

  In terms of the signal flow diagram of~\cref{fig:signalflowgraph}, the following computations are related to reducing the squeezer subgraph of part C. For that case, there are two cavity inputs and one cavity output. The primary input and output are where the parametric gain is applied, and there is an additional input for the internal loss of the squeezer cavity, which we apply only on one side of the crystal. There is technically one additional physically-necessary output for the loss, but it is ignored for the model.

  The system of equations is established by the graph on the right of \cref{fig:signalflowgraph}-C.
\subsection{Calculating $G_{\text{int}}^{\text{f}}$ and $G_{\text{int}}^{\text{b}}$}
The single-pass squeezing (power) gain $\text{G}_\text{sqz} = e^{-2z}$ through the nonlinear crystal can be used to calculate an effective gain for the OPO $\Gfint$. We calculate this gain following the formalism in Ganapathy et. al ~\cite{GanapathyPRD22ProbingSqueezing}. The action of squeezing mixes the positive and negative sidebands which manifests as an amplitude attenuation or amplification in the quadrature basis:
\begin{align}
\label{eq:Squeezer}
    \hat{S^\dagger}(z) (\hat{X}_{1,\text{in}} + i \hat{X}_{2,\text{in}})\mathbf{S}(z) &= e^{-z} \hat{X}_{1,\text{out}} + i e^{z} \hat{X}_{2,\text{out}}
  &&&
      \hat{X}_{1,\text{in}} &\equiv \frac{1}{\sqrt{2}}\left(\hat{a}_{\text{z,in}} + i\hat{a}^\dagger_{\text{z,in}}  \right)
  &&&
      \hat{X}_{2,\text{in}} &\equiv \frac{1}{\sqrt{2}}\left(\hat{a}_{\text{z,in}} - i\hat{a}^\dagger_{\text{z,in}}  \right)
\end{align}
$\hat{S}(z)$ represents the squeezing operator and $\hat{X_1}$ and $\hat{X_2}$ are the real and imaginary components of the electric field amplitude which represent the two quadrature operators~\cite{WallsQO08QuantisationElectromagnetic}. In a nearly lossless bow-tie OPO, where all mirrors excluding the coupling mirror are perfectly reflective, the round trip phase for a sideband of frequency $\Omega$ above the carrier is given by $e^{-i\frac{\Omega L_z}{c}}$, where $L_z$ the length around the cavity.

When the carrier (pump field) and the signal are injected via the squeezer cavity input mirror, a squeezed coherent field is obtained on reflection at squeezer cavity input mirror. The effective gain for the squeezed quadrature can be calculated by considering both the reflected field and multiple round trips within the bow-tie cavity. 
\begin{align}
\hat{X}_{\text{1,out}} &= \sqrt{\text{G}_{\text{int}}^{\text{f,b}}}\hat{X}_{\text{1,in}} &&&
\hat{X}_{\text{2,out}} &= \frac{1}{\sqrt{\text{G}_{\text{int}}^{\text{f,b}}}}\hat{X}_{\text{2,in}} &&&
  \sqrt{\text{G}_{\text{int}}^{\text{f,b}}} = e^{\mp Z}
   &= -r_1 + \frac{t_1^2(e^{-i\frac{\Omega L_z}{c}} e^{\mp z})}{1-r_1e^{-i\frac{\Omega L_z}{c}}e^{\mp z}}
\end{align}
If we let the cavity be operated at resonance for the squeezed vacuum carrier frequency, the squeezed quadrature gain reduces to~\cite{GanapathyPRD22ProbingSqueezing}:
\begin{equation}
   \sqrt{\text{G}_{\text{int}}^{\text{f,b}}} = e^{\mp Z} = \frac{e^{\mp z} - r_1}{1-r_1 e^{\mp z}} = \frac{\sqrt{\text{G}^{\text{f,b}}_\text{sqz}} - r_1}{1-r_1\sqrt{\text{G}^{\text{f,b}}_\text{sqz}}}
\end{equation}

\subsection{Calculating $\epsilon_{\rm sqz}\mapsto(\epsseII,\epsseIII)$}\label{sup:lossmap}

We find the transfer function of the vacuum from the loss port inside the squeezer to just outside the squeezer $H_{ \epssqz\to \epsilon_{\rm se(\#)}}$. We can then model this loss inside the squeezer cavity by setting $\epsilon_{se(\#))} = |H_{ \epssqz\to \epsilon_{\rm se(\#)}}|^2\epssqz$. We show the derivation for $\epsseII,\epsseIII$, as these are the only ports that contribute to the shot noise in our proposed scheme. The derivation is analogous for the radiation pressure noise.

We consider a four-mirror bow tie squeezer cavity as described above (with input coupler labelled mirror 1, with reflectivity and transmissivity $r_1+t_1=1$, mirror 2 with  $r_2=1-t_2-\sqrt{\epssqz}$ and mirrors 3 and 4 with $r_{3,4}=1,\,t_{3,4}=0$ (where number labels are ascending for the forward propagation direction), For reasons described above, the loss $\epssqz$ is assumed to be incurred on mirror 2 of the squeezer cavity.

\noindent\textbf{Forward direction: $\epsilon_{\mathrm{sqz}} \to \epsilon_{\rm se2}$.}

Given the transfer function of a cavity with two imperfect mirrors, we can write
\begin{equation}
\sqrt{\epsilon_{se2}}
=|H_{ \epssqz\to \epsilon_{\rm se(2)}}|\sqrt{\epssqz}
=
\frac{t_1\, {\sqrt{\text{G}^{\text{f}}_\text{sqz}}\, r_2}}{1 - r_1\, r_2\, \sqrt{\text{G}^{\text{f}}_\text{sqz}}}\; \sqrt{\epsilon_{\mathrm{sqz}}} .
\label{eq:fwd-cavity}
\end{equation}
Now we approximate 
\begin{equation}
r_2 \approx 1, \qquad
\sqrt{\text{G}^{\text{f}}_\text{sqz}} = {\sqrt{\text{G}^{\text{f,th}}_\text{sqz}}} \approx r_1 \sqrt{1 - \epsilon_{\mathrm{sqz}}} ,
\label{eq:fwd-threshold}
\end{equation}
i.e. keeping only terms first order in loss and setting the squeezing to be at the OPO threshold. Inserting~\eqref{eq:fwd-threshold} into~\eqref{eq:fwd-cavity}:
\begin{equation}
\sqrt{\epsilon_{se2}}
\;=\;
\frac{t_1\, r_1\, \sqrt{1 - \epsilon_{\mathrm{sqz}}}}
     {1 - r_1^{\,2}\, \sqrt{1 - \epsilon_{\mathrm{sqz}}}}
\; \sqrt{\epsilon_{\mathrm{sqz}}} .
\label{eq:fwd-amp}
\end{equation}
Squaring both sides yields the power relation:
\begin{equation}
\epsilon_{se2}
\;=\;
\frac{t_1^{\,2}\, r_1^{\,2}\, (1 - \epsilon_{\mathrm{sqz}})}
     {\bigl(1 - r_1^{\,2}\, \sqrt{1 - \epsilon_{\mathrm{sqz}}}\bigr)^{2}}
\; \epsilon_{\mathrm{sqz}} .
\label{eq:fwd-power}
\end{equation}
For $\epsilon_{\mathrm{sqz}} \ll 1$, we expand
$\sqrt{1 - \epsilon_{\mathrm{sqz}}} \approx 1$ and use $r_1^{\,2} + t_1^{\,2} = 1$, i.e.\ $1 - r_1^{\,2} = t_1^{\,2}$:
\begin{equation}
\epsilon_{se2}
\;\approx\;
\frac{t_1^{\,2}\, r_1^{\,2}}{(1 - r_1^{\,2})^{2}}\; \epsilon_{\mathrm{sqz}}
\;=\;
\frac{t_1^{\,2}\, r_1^{\,2}}{t_1^{\,4}}\; \epsilon_{\mathrm{sqz}}
\;=\;
\frac{r_1^{\,2}}{t_1^{\,2}}\; \epsilon_{\mathrm{sqz}}
\;=\;
\frac{R_1}{T_1}\; \epsilon_{\mathrm{sqz}} ,
\label{eq:fwd-smallloss}
\end{equation}
where $R_1 \equiv r_1^{\,2}$ and $T_1 \equiv t_1^{\,2}$ are the power
reflectivity and transmissivity of the input coupler.

\bigskip

\noindent\textbf{Backward direction:
$\epsilon_{\mathrm{sqz}} \to \epsilon_{\rm se3}$.}
The derivation proceeds analogously as for the forward direction and we find:
\begin{equation}
\epsilon_{\rm se3}
\;=\;
\frac{t_1^{\,2}\, r_1^{\,2}\, (1 - \epsilon_{\mathrm{sqz}})}
     {\bigl(1 - r_1^{\,2}\, \sqrt{1 - \epsilon_{\mathrm{sqz}}}\bigr)^{2}}
\; \epsilon_{\mathrm{sqz}}
\;=\; \epsilon_{\rm se2} .
\label{eq:bwd-power}
\end{equation}
In the same small-loss limit,
\begin{equation}
\epsilon_{\rm se3} \;\approx\; \frac{R_1}{T_1}\, \epsilon_{\mathrm{sqz}} .
\label{eq:bwd-smallloss}
\end{equation}

\widetext

\end{document}